\newif\ifemulate
\emulatetrue

\ifemulate
	\documentclass[iop,revtex]{emulateapj-rtx4}
	\slugcomment{Accepted for Publication in \apj}
	\lefthead{Rudnick et al.}
	\righthead{Halo mass scale for gas removal}
\else
	\documentclass[12pt, preprint]{aastex}
\fi

\usepackage{graphicx}
\usepackage{lscape}
\usepackage{natbib}
\usepackage{url}

\newcommand\hda{H$\delta_{abs}$}
\newcommand\hga{H$\gamma_{abs}$}
\newcommand\hdga{$\langle {\rm H}\delta_{abs}{\rm H}\gamma_{abs}\rangle$}
\newcommand\ewhbe{EW(H$\beta_{em}$)}
\newcommand\ewhda{EW(H$\delta_{abs}$)}
\newcommand\ewhga{EW(H$\gamma_{abs}$)}
\newcommand\ewoii{EW([\ion{O}{2}])}
\newcommand\ewoiii{EW([\ion{O}{3}])}
\newcommand\oii{[\ion{O}{2}]}
\newcommand\oiii{[\ion{O}{3}]}
\newcommand\dn{$D_n$(4000)}

\newcommand\ang{${\rm \AA}$}

\newcommand\mlstar{${ M_{\star}}/{\rm L}$}

\newcommand{\mlstarlamav}[1]{$\langle {\cal M_{\star}}/{\rm L}_{V}\rangle$}

\newcommand\msol{${\cal M_{\odot}}$}
\newcommand\lsol{$L_{\odot}$}

\newcommand{\mstar}{${\cal M_{\star}}$}

\newcommand\lir{$L_{IR}$}
\newcommand\atlas{ATLAS$^{\rm3D}$}

\def\gtsima{$\; \buildrel > \over \sim \;$}
\def\gsim{\lower.7ex\hbox{\gtsima}}
\def\ltsima{$\; \buildrel < \over \sim \;$}
\def\lsim{\lower.7ex\hbox{\ltsima}}

\renewcommand\micron{$\mu$m}

\def\apj{ApJ}

\begin{document}
\title{Determining the Halo Mass Scale where Galaxies Lose Their Gas.\footnote{\lowercase{Based on observations obtained at the
      European Southern Observatory using the ESO Very Large Telescope
      on Cerro Paranal through ESO program 166.A-0162.}}}

\author{Gregory Rudnick\altaffilmark{1,2}, 
Pascale Jablonka\altaffilmark{3,4}, 
John Moustakas\altaffilmark{5}, 
Alfonso Arag\'on-Salamanca\altaffilmark{6}, 
Dennis Zaritsky\altaffilmark{7}, 
Yara L. Jaff\'e\altaffilmark{8,9}, 
Gabriella De Lucia\altaffilmark{10}, 
Vandana Desai\altaffilmark{11}, 
Claire Halliday\altaffilmark{12}, 
Dennis Just\altaffilmark{7,13},
Bo Milvang-Jensen\altaffilmark{14}, 
Bianca Poggianti\altaffilmark{15}}

\altaffiltext{1}{The University of Kansas, Department of Physics and Astronomy, Malott room 1082, 1251 Wescoe Hall Drive, Lawrence, KS, 66045, USA; \texttt{grudnick@ku.edu}}
\altaffiltext{2}{Alexander von Humboldt Fellow at the Max-Planck-Institute f\"{u}r Astronomy, Heidelberg, Germany}
\altaffiltext{3}{Institut de Physique, Laboratoire d' Astrophysique, Ecole Polytechnique F\'ed\'erale de Lausanne (EPFL), Observatoire de Sauverny, CH-1290 Versoix, Switzerland}
\altaffiltext{4}{GEPI, Observatoire de Paris, CNRS, Universit\'e Paris Diderot, F-92125 Meudon Cedex, France}
\altaffiltext{5}{Department of Physics \& Astronomy, Siena College, 515 Loudon Road, Loudonville, NY 12211, USA}
\altaffiltext{6}{School of Physics and Astronomy, The University of Nottingham, University Park, Nottingham NG7 2RD, UK}
\altaffiltext{7}{University of Arizona, 933 N. Cherry Ave, Tucson, AZ 85721, USA}
\altaffiltext{8}{Department of Astronomy, Universidad de Concepci\'on, Casilla 160-C, Concepci\'on, Chile}
\altaffiltext{9}{European Southern Observatory, Alonso de Cordova 3107, Vitacura, Santiago, Chile}
\altaffiltext{10}{INAF-Osservatorio Astronomico di Trieste, Via Tiepolo 11, I-34131 Trieste, Italy}
\altaffiltext{11}{Spitzer Science Center, California Institute of Technology, MS 220-6, Pasadena, CA 91125, USA}
\altaffiltext{12}{23, rue d'Yerres, 91230 Montgeron, France}
\altaffiltext{13}{Department of Astronomy \& Astrophysics, University of Toronto, 50 St George Street, Toronto, Ontario M5S 3H4, Canada}
\altaffiltext{14}{Dark Cosmology Centre, Niels Bohr Institute, University of Copenhagen, Juliane Maries Vej 30, DK-2100 Copenhagen, Denmark}
\altaffiltext{15}{INAF-Osservatorio Astronomico di Padova, Vicolo dell Osservatorio 5, I-35122 Padova, Italy}

\begin{abstract}

A major question in galaxy formation is how the gas supply that fuels
activity in galaxies is modulated by their environment.  We use
spectroscopy of a set of well characterized clusters and groups at
$0.4<z<0.8$ from the ESO Distant Cluster Survey (EDisCS) and compare
it to identically selected field galaxies.  Our spectroscopy allows us
to isolate galaxies that are dominated by old stellar populations.
Here we study a stellar-mass limited sample ($\log(M_*/M_\odot)>10.4$)
of these old galaxies with weak \oii\ emission. We use line ratios and
compare to studies of local early type galaxies to conclude that this
gas is likely excited by post-AGB stars and hence represents a diffuse
gas component in the galaxies.  For cluster and group galaxies the fraction with \ewoii$>5$\ang\ is
$f_{[OII]}=0.08^{+0.03}_{-0.02}$ and $f_{[OII]}=0.06^{+0.07}_{-0.04}$
respectively.  For field galaxies we find
$f_{[OII]}=0.27^{+0.07}_{-0.06}$, representing a 2.8$\sigma$
difference between the \oii\ fractions for old galaxies between the
different environments.  We conclude that a population of old galaxies
in all environments has ionized gas that likely stems from stellar
mass loss.  In the field galaxies also experience gas accretion from the cosmic web and in groups and
clusters these galaxies have had their gas accretion shut off by their
environment.  Additionally, galaxies with emission preferentially
avoid the virialized region of the cluster in position-velocity space.  We discuss the implications of our results, among which is that gas accretion
shutoff is likely effective at group halo masses (log~${\cal
  M}/$\msol$>12.8$) and that there are likely multiple gas removal processes
happening in dense environments.

\end{abstract}
\section{Introduction}
\label{Sec:intro}

One of the longest standing problems in galaxy evolution is how star formation in galaxies is quenched.  We have known for over 50 years that there is a significant population of galaxies with uniformly red colors that occupy a tight sequence in color and magnitude \citep[e.g.][]{deVauc61,Visvanathan77}. \citet{Couch83}, \citet{Butcher84}, \citet{Bower92a} and \citet{Bower92b} interpreted this tight sequence as resulting from uniformly old stellar ages for these ``red sequence" galaxies.  The tight scatter in color of the red sequence in turn implies that galaxies must stay passive for extended periods \citep[e.g.][]{AragonSalamanca93,Bower98,Kodama98}.  Subsequently, it was demonstrated that the distribution of galaxy star formation rates (SFRs) and star formation histories (SFHs) is approximately bimodal, with galaxies either forming stars or being passive \citep[e.g.][]{Strateva01,Kauffmann03}, although this bimodality might be less pronounced for the most massive galaxies \citep[e.g.][]{Salim07}.  Lookback studies have shown that the amount of stellar mass in the star-forming population remains relatively constant at $z<1.5$, while the mass in the passive population increases over the same epoch \citep[e.g.][]{Bell04,Blanton06,Brammer11}.  This is as would be expected if star-forming galaxies were quenched and added to the passive population progressively over time.  However, despite the ample evidence for galaxy quenching, it is not clear if this quenching is rapid \citep[e.g.][]{Bell04} or slow \citep[e.g.][]{Schawinski14}.

At about the same time, it was also realized that the fraction of galaxies that are passive depends on environment.  In the local rich clusters such as Coma, the passive population completely dominates \citep{Terlevich01}, but even in less extreme environments it is clear that denser environments host a larger fraction of passive galaxies \citep{Hogg04,Blanton09}.  One possible explanation for these observed correlations is that there is are physical processes that occur in dense environments that suppress star formation in galaxies.  The color-density relation holds to at least $z=2$ \citep{Cooper06,Gerke07,Peng10,Cooper10b,Quadri12}, and if this is indicative of an active suppression of star formation in dense environments, indicates that these processes may have been important  over a majority of cosmic time.  However, an alternate explanation for the SFR-density relation is that galaxies in dense environments are simply older than those in lower density regions, presumably echoing an earlier formation epoch \citep[e.g.][]{Gao05,Tonnesen15}.

It is difficult to disentangle these two possibilities and much work has been done to isolate the signatures of specific transformative processes.   On the observational side, there is mounting evidence that whatever shuts off star formation in galaxies entering dense environments must precede a transformation in the morphology.  For example, the rapid build-up in the red sequence luminosity function at $z<0.8$ \citep[][but see also Andreon 2008, Crawford et al. 2009, \& De Propris et al. 2013]{DeLucia04,Tanaka05,Stott07,DeLucia07a,Gilbank08,Rudnick12} seems to precede the buildup in the S0 population that becomes significant at $z<0.5$ \citep{Desai07b}.  This sequencing of gas and morphological processes is also supported by the lack of blue S0 galaxies \citep{Jaffe11a} in clusters and the lack of UV-emission in S0s found in a merger of galaxy groups \citep{Just11}, both of which indicate that galaxies change their morphology after their star formation is suppressed.  Indeed, in \citet{Gallazzi09} and \citet{Wolf09} it was found that there is a population of spiral galaxies in the A901/902 supercluster that have spiral morphology but suppressed star formation.  A similar population of galaxies is even found in intermediate redshift clusters \citep{Cantale16} and in the field \citep{Bundy10}.

The quenching of star formation without a commensurate change in morphology argues strongly for processes which suppress star formation by the depletion of gas in galaxies.  There are multiple theoretical mechanisms for the gas to be affected in galaxies.  Strangulation or starvation, first proposed by \citet{Larson80} is now understood to encompass the broad family of processes in which galaxies decouple from the gas that flows into them from the intergalactic medium (IGM) and thus use up their internal gas supply on an extended timescale of a few Gyr.  This process seems to match the observed increase with time in the quenched fraction of galaxies in dense environments \citep[e.g.][]{McGee11,DeLucia12a,Taranu14,Haines15}.  This mechanism should be effective all the way down to at least group scales as satellite galaxies in those haloes should be decoupled from their gas flows \citep{Kawata08}.  Indeed, observations of galaxies in groups find them to be HI deficient with respect to matched galaxies in the field \citep{Catinella13}.  

A modification of the starvation scenario discussed by various authors has been the role of winds in speeding up the depletion of gas once accretion is shut off \citep[e.g.][]{Weinmann06,Wang07}.  These winds are known to be ubiquitous in star-forming galaxies \citep[e.g.][]{Weiner09,Rubin10,Rubin14} but with an uncertain mass-loading.  Recently, \citet{McGee14} found that winds with reasonable mass-loading factors could, when coupled with a long-delay time upon becoming a satellite \citep{Wetzel13}, explains both the long timescales for transformation to happen for infalling satellite galaxies and the rapid shutoff in star formation once the delay time has passed.  This wind-driven consumption was termed by \citet{McGee14} as ``overconsumption" and is able to explain the observed weak dependence of the SFR of SF galaxies on environment, which implies that SF must be shut off quickly \citep{Peng10,McGee11}.

An alternate process for the suppression of star formation is the stripping of the cold gas via ram pressure effects \citep{Gunn72} which results from a galaxy passing through the dense intracluster medium (ICM).  Theoretical investigations have shown that this process can remove significant amounts of both diffuse and cold gas from within the optical radius of galaxies in rich cluster environments \citep{Quilis00,Roediger07,Roediger08,Tonnesen09}.  Observational investigations have shown both dramatic stripping of the gas content of galaxies in cluster environments \citep{Kenney99,Kenney04,Cortese11,Pappalardo12,Fumagalli14,Jachym14,Boselli14,Jaffe15} and statistical evidence from rotation curve asymmetries in cluster galaxies with normal morphology that ram pressure stripping is likely important in those environments \citep{Bosch13}.  

Contrary to common conceptions, ram pressure stripping need not be fast ($\sim 100$Myr) as the fast timescales inferred from theoretical investigations \citep{Quilis00} are only valid if galaxies are inserted at high speeds into a dense ICM.  Galaxies falling into the cluster from larger radii instead will experience ram pressure stripping on a longer timescale more akin to a crossing time, as the pressure builds up during the infall process as a result of the increasing velocity and ICM density.  The actual modification to the gas of a galaxy suffering ram pressure stripping depends sensitively on the orbit of the galaxy at infall as galaxies with different orbits will spend different times in the region of the cluster with the highest ram pressure \citep[e.g.][]{Brueggen08,Jaffe15}.

Despite much work on understanding the effects of environment on galaxy gas, the relative roles of starvation and ram pressure stripping have proven difficult to disentangle as the timescales may not be so different as previously thought, and because few observations trace a large dynamic range in density and contain a significant number of high density systems.  It is clear that there is a population of star-forming galaxies in clusters that exhibit slightly suppressed SFRs \citep{Vulcani10,Haines13}, which may indicate a slow quenching process.  However, there is still a significant degree of tension between the quenching timescales inferred by the above physical mechanisms and the very long quenching timescales needed to reproduce the evolution in the galaxy passive fractions over cosmic time \citep{McGee11,DeLucia12a,Wetzel13}.    It is therefore necessary to examine both of these processes in more detail and to determine which of the processes are required by the observations, which are ruled out, which are merely allowed, and over what halo mass ranges these different conclusions hold.

An additional question that needs to be resolved is how to keep dead (or passive) galaxies dead.  Galaxies replenish their gas through mass loss, with a \citet{Chabrier03} or \citet{Kroupa01} initial mass function (IMF) returning $\sim 40(50)\%$ of a simple stellar population's stellar mass to gas within 1(13)Gyr.  It is not clear how to keep that gas from forming stars.  This gas is observed to exist in local galaxies, as studies have shown that emission lines from ionized gas (e.g. [\ion{O}{2}], H$\alpha$) in passive galaxies are common, may be supplied by mass loss, and may be predominantly heated by existing stellar populations \citep{Yan06,Sarzi06,Sarzi10,Yan12,Singh13,Belfiore16}.  There are some indications as well from integral field spectroscopic identification of distinct kinematic components that this gas can be supplied by external accretion and that external accretion is not a significant supply of replenished gas in Virgo cluster galaxies \citep{Davis11}.  Despite the insight granted by their 3D (spatial$+$kinematic) data, these local studies are limited to only a single cluster and do not adequately probe the group environment.  Thus, the effect of environment on this supply of gas is still not well understood.

To better understand the mechanisms through which the gas supply is modulated in galaxies, we must first find a population of galaxies which have a specific signature of gas depletion/removal.  We must then examine how this population changes as a function of halo mass, across a broad range of stellar mass, and also how the population changes as a function of position and velocity with respect to the host halo.   That is the purpose of this paper.

Studying galaxies in intermediate redshift clusters is a promising place to examine the gas mechanisms at play when clusters were growing rapidly and when galaxy quenching was proceeding at a rapid pace.  While studies in the local universe, for example using SDSS data, have a plethora of line indices to characterize the gas, SDSS is subject to significant aperture biases \citep[e.g.][]{Brinchmann04,Labbe07} that complicate the interpretation of the line emission in terms of the global properties of the galaxies.  On the other hand, slit spectra of intermediate redshift galaxies capture a majority of the galaxy light, making them more straightforward probes of integrated galaxy properties.  While integral field spectroscopic surveys like SAURON \citep{deZeeuw02}, ATLAS$^\mathrm{3D}$ \citep{Cappellari11}, and CALIFA \citep{Singh13} probe spatially integrated emission processes in the nearby Universe, and especially the case of emission in passive galaxies \citep[e.g.][]{Sarzi06}, they only contained one dense environment, i.e. the Virgo cluster, making it difficult for those surveys to draw significant conclusions about the effect of environment.  New new integral field surveys like SDSS-Manga \citep{Bundy15} and SAMI \citep{Bryant15} are starting to probe environmental dependence of gas properties.  However, there is now evidence that the quenching mechanism may have been different at higher redshift, due to the different density of the ICM and the higher gas fractions of galaxies \citep[e.g.][]{Balogh16}.  Therefore, if we want to
understand gas accretion shutoff processes at intermediate redshift, which are responsible for the quenching of many passive galaxies in today's clusters,
it is best to probe intermediate redshift galaxies directly.

We employ spectroscopic observations of a large sample of field, group, and cluster galaxies at $0.4<z<0.8$ collected as part of the ESO Distant Cluster Survey \citep[EDisCS;][]{White05}.  Our spectroscopic observations of \hda\ and \hga\ Balmer absorption lines, the 4000\ang\ break, and various emission lines allow us to determine the recent SFH and current SFR of our galaxies in a stellar mass selected sample \citep{Barger96,Rudnick00,Kauffmann03}.  We isolate a population of galaxies dominated by older stellar populations, yet having emission lines and then determine what their prevalence and characteristics tell us about gas stripping and starvation properties in dense environments.  We examine many possibilities to explain the observed results, and conclude on what possibilities are required by the observations and which are merely allowed.

The outline of our paper is as follows.  In \S\ref{Sec:data} we discuss our sample, photometric and spectroscopic measurements, and stellar mass determination.  In \S\ref{Sec:sel} we discuss our use of spectral indices to isolate galaxies with different relative stellar ages and we discuss their broad-band spectral properties.  In \S\ref{Sec:env_dep} we present the environmental dependence of our older galaxies with emission lines, including their distribution in velocity and position in our clusters.  In \S\ref{Sec:disc} we discuss the origin of the observed environmental differences and use a comprehensive decision tree to decide on what is allowed and required by our observations.  We summarize and conclude in \S\ref{Sec:summ}.

Throughout we assume ``concordance'' $\Lambda$-dominated cosmology
with
$\Omega_\mathrm{M}=0.3,~\Omega_{\Lambda}=0.7,~\mathrm{and~H_o}=70~{\rm
  h_{70}~km~s^{-1}~Mpc^{-1}}$ unless explicitly stated otherwise.  All
magnitudes are quoted in the AB system \citep{Oke74}.

\section{Data}
\label{Sec:data}

\subsection{Sample Definition, Photometry, and Morphologies}

Our data are in fields observed by the ESO Distant Cluster Survey
(EDisCS).  Each field was selected to have a primary optically
selected cluster.  The exact selection of the parent EDisCS fields and
the main clusters is described in detail in \citet{White05}.  In some
fields there are additional clusters and groups serendipitously
discovered along the line of sight.  Each field was observed with
$BVIK_s$, $BVIJK_s$, or $VRIJK_s$ photometry depending on the initial
redshift estimate.  The optical data were obtained with VLT/FORS2 and the NIR data with NTT/SOFI.  Vega-to-AB magnitude conversions were computed by integrating the spectrum of Vega over our filter curves.  

As part of EDisCS we have obtained 30-50
spectra of members per main cluster and significant numbers in the
projected groups and clusters along the line of sight as well as many
field galaxies at similar redshifts to the main structures
\citep{Halliday04,Milvang08}.  For objects that were observed multiple times, we replaced the multiple occurrences with a single measurement that is the average of the individual measurements.

We restrict our sample of groups and field galaxies to those within
$\Delta z=\pm 0.2$ of the main cluster redshift.  It was shown in \citet{Halliday04} and \citet{Milvang08} that the EDisCS spectroscopic target selection, which combined an $I$-band magnitude cut and loose photometric redshift cuts \citep{Pello09}, missed an estimated 3\% of the galaxies at the targeted cluster redshift compared to a strictly $I$-band selected sample.  Our spectroscopy within $\Delta z=\pm 0.2$ of the main cluster redshift is therefore unbiased. Following
\citet{Poggianti09} we classify clusters as those systems with a
velocity dispersion $\sigma>400$km/s and groups as those with at least
8 spectroscopic members and $160$km/s$<\sigma\leq 400$km/s.  We
further require our systems to have adequate exposures in all EDisCS
bands to allow for the computation of stellar masses (see below).
This selection results in 17 clusters and 7 groups.  Two groups were
excluded because they lie outside the area of the EDisCS images with
NIR photometry and therefore their stellar masses could not be
computed robustly because our observations did not probe to red rest-frame optical wavelengths.  The properties of our sample of structures are listed in
Table~\ref{clustdat_tab}.

As galaxies evolve with redshift it is important to control for differences in the redshift distribution of the galaxy samples in different environments.  We test for this by comparing the redshift distributions for the final sample of galaxies in each of our three subsamples (cluster, group, and field; see \S\ref{Sec:mass_compl} for final sample size).  The median and 68\% limits of the redshit distribution for the three samples are $z_{\rm med,clust}=0.60_{-0.14}^{+0.19}$, $z_{\rm med,group}=0.58_{-0.18}^{+0.15}$, and $z_{\rm med,field}=0.63_{-0.14}^{+0.13}$.  Therefore the distributions for all samples are identical and our results will not be affected by different redshift distributions.

Using the velocity dispersion and redshift we derive the halo mass
for our structures using Equation 10 from \citet{Finn05}.  Our clusters occupy the halo mass range of
$10^{13.8}-10^{15.2}$\msol\ and our groups correspond to
$10^{12.8}-10^{13.8}$\msol.  It is important to note that the halo masses
and velocity dispersions for our groups, especially the poorer groups,
are quite uncertain.  Membership in these structures is based on a
3$\sigma$ cut in the velocity dispersion of the system, where $\sigma$
is computed iteratively using the biweight scale estimator
\citep{Beers90,Halliday04,Milvang08}.  In this paper we only study the
spectroscopic sample of galaxies.  Our
``field'' sample consists of true field galaxies in the sense that they do not
belong to any structure with $\sigma>160$km/s.

\ifemulate
	\begin{deluxetable}{clll}
\else
	\begin{deluxetable}{clll}
\fi
\tablecaption{Cluster and Group Data}
\tablewidth{0pt}
\tablehead{\colhead{system} & \colhead{$z$} & \colhead{$\sigma$\tablenotemark{a}} & \colhead{$N_{samp}$\tablenotemark{b}}\\
\colhead{} & \colhead{} &  \colhead{[km/s]}  & \colhead{}}
\startdata
\\
& Clusters\\
\hline
\\
cl1018.8-1211 & 0.4734 & $486^{+59}_{-63}$ & 17\\
cl1040.7-1155 & 0.7043 & $418^{+55}_{-46}$ & 10\\
cl1054.4-1146 & 0.6972 & $589^{+78}_{-70}$ & 26\\
cl1054.7-1245 & 0.7498 & $504^{+113}_{-65}$ & 21\\
cl1059.2-1253 & 0.4564 & $510^{+52}_{-56}$ & 26\\
cl1138.2-1133 & 0.4796 & $732^{+72}_{-76}$ & 9\\
cl1138.2-1133a & 0.4548 & $542^{+63}_{-71}$ & 5\\
cl1202.7-1224 & 0.4240 & $518^{+92}_{-104}$ & 9\\
cl1216.8-1201 & 0.7943 & $1018^{+73}_{-77}$ & 42\\
cl1227.9-1138 & 0.6357 & $574^{+72}_{-75}$ & 13\\
cl1232.5-1250 & 0.5414 & $1080^{+119}_{-89}$ & 9\\
cl1301.7-1139 & 0.4828 & $687^{+81}_{-86}$ & 18\\
cl1353.0-1137 & 0.5882 & $666^{+136}_{-139}$ & 13\\
cl1354.2-1230 & 0.7620 & $648^{+105}_{-110}$ & 11\\
cl1354.2-1230a & 0.5952 & $433^{+95}_{-104}$ & 6\\
cl1411.1-1148 & 0.5195 & $710^{+125}_{-133}$ & 13\\\\
\hline
\\
& Groups\\
\hline
\\
cl1037.9-1243 & 0.5783 & $319^{+53}_{-52}$ & 7\\
cl1040.7-1155a & 0.6316 & $179^{+40}_{-26}$ & 6\\
cl1040.7-1155b & 0.7798 & $259^{+91}_{-52}$ & 2\\
cl1054.4-1146a & 0.6130 & $227^{+72}_{-28}$ & 5\\
cl1054.7-1245a & 0.7305 & $182^{+58}_{-69}$ & 9\\
cl1227.9-1138a & 0.5826 & $341^{+42}_{-46}$ & 1\\
cl1301.7-1139a & 0.3969 & $391^{+63}_{-69}$ & 10\\
cl1420.3-1236 & 0.4962 & $218^{+43}_{-50}$ & 18\\
\enddata
\label{clustdat_tab}
\tablenotetext{a}{This was computed using the full EDisCS member sample, which is larger than the sample used for the analysis in this paper.}
\tablenotetext{b}{The number of galaxies meeting our stellar mass and quality cuts.}
\ifemulate
	\end{deluxetable}
\else
	\end{deluxetable}
\fi

Rest-frame luminosities were derived for each galaxy from the full spectral energy distribution (SED)
using the technique described in \citet{Rudnick09} and adopting the
spectroscopic redshift.  \footnote{Two of our systems, CL1138.2-1133
  and CL1138.2-1133a, do not have observed $B$-band photometry and
  with $z=0.48$ and $0.45$ respectively lie at redshifts slightly below
  that where the rest-frame $U$-band is purely interpolated from our
  photometry ($z=0.53$).  The extrapolation to rest-frame $U$-band,
  however, is small, with a minimal effect on the actual colors.  The
  rest-frame luminosities in this paper are illustrative only and are
  not crucial to any part of our analysis.  We have verified that the
  color-mass distribution of these clusters is similar to that of the
  others and include them in our sample.}

We use visual morphologies derived from \textit{Hubble Space
  Telescope} (HST) F814W imaging of 10 of EDisCS fields
\citep{Desai07a}.  This observed filter corresponds to rest-frame $B$-
or $V$-band imaging of our systems depending on the redshift of the
sample.  The morphologies were derived by visual classifications by
multiple team members \citep{Desai07a}.  The fields
with HST morphology contain 11 of our sample clusters and 5 of our
groups.

\subsection{Stellar Masses}

Stellar masses were derived using the iSEDfit code presented in \citet{Moustakas13}.  In brief summary, we fit the observed photometry
using a Bayesian technique that assumes exponentially declining or
constant SFHs with superimposed random bursts.  We use the Flexible
Stellar Population Synthesis (FSPS) library from \citet{Conroy09} and
\citet{Conroy10} with a \citet{Chabrier03} initial mass
function (IMF) and a \citet{Charlot00} attenuation curve.  The IMF extends from $0.1-100$~\msol.  The masses
derived with the FSPS library are less than 0.1~dex different than
when using the \citet{Bruzual03} library.  Our choice of attenuation
curve also results in a difference in the masses of less than 0.1~dex
when compared to the \citet{Calzetti00} curve.  The inclusion of
bursts in the SFH as opposed to a smooth SFH results in a difference
of less than 0.1~dex for most galaxies, with up to an 0.3~dex
difference for individual objects.

\begin{figure*}
\plotone{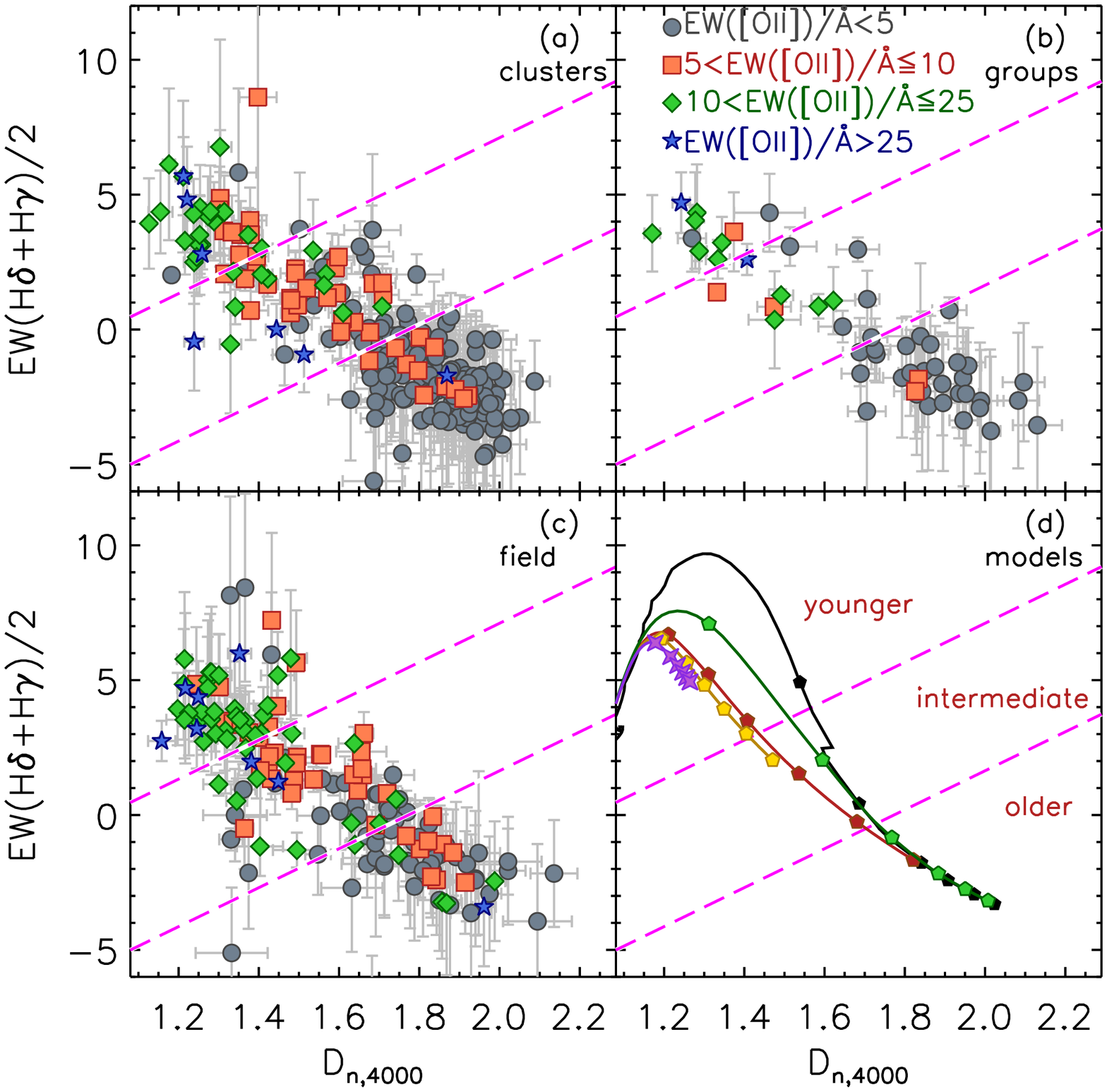}
\caption {A plot of the average of the equivalent width of \ewhda\ and
  \ewhga\ vs \dn\ for cluster (a), group (b), and field (c) galaxies.
  All emission and absorption features have been decomposed as
  described in the text.  Only galaxies with log(\mstar/\msol$)>10.4$ are
  plotted.  The color of the points indicates \ewoii\ as described in
  the legend.  The tracks in (d) are the \citet{Bruzual03} predictions
  for the evolution of a stellar population formed in a single burst
  (upper black curve), with a constant SFR (purple curve), and with three exponentially declining SFHs with a 300~Myr timescale (upper green curve), a
  1~Gyr timescale (middle red curve), and a 2~Gyr timescale (lower
  yellow curve).  The solid lines represent models having solar
  metallicity.  All models extend to an age of 6~Gyr, which
  corresponds roughly to a start of star formation at $z=2.5$ for a
  galaxy observed at the lower end of the EDisCS redshift range
  $z=0.4$.  The points mark 1~Gyr intervals increasing to the lower
  right.  Only old simple stellar populations or exponentially
  declining SFHs after many e-foldings can cross the lower dashed
  magenta line.  As described in the text we therefore divide our
  galaxies by the relative age of the stellar population using the
  magenta lines, which have a slope of 7.2.}
\label{Fig:d4hd}
\end{figure*}

\subsection{Spectral Indices}

Much of our analysis relies on the ability to measure absorption
features and weak emission lines from our spectra.  The EDisCs spectra are well calibrated \citep{Halliday04,Milvang08} and also allow the computation of line ratios.  As
we demonstrate in \S\ref{Sec:col}, the indices we employ are far less
susceptible to reddening than rest-frame optical colors and give a
more clear mapping from the indices to a SFH.  Specifically, we are
interested in the age-sensitive Balmer absorption (\hda\ and \hga) and
4000\AA\ break (\dn) stellar continuum features, as well as the strong
nebular emission lines [\ion{O}{2}]~$\lambda3727$, [\ion{O}{3}]~$\lambda\lambda4959,5007$, and $H\beta$.  The \dn\ index first introduced by \citet{Balogh99} has a shorter wavelength baseline than the traditional $D$(4000) \citep{Bruzual83} and is therefore less susceptible to dust extinction.  Our desire to separate out the emission and absorption components presents
complications as the emission lines can fill in the absorption
features and because the emission line measurements are difficult in
regions of the spectra with many absorption features
\citep[e.g.][]{Rudnick00}.  We address these complications by
decomposing the emission and absorption using the techniques described
in detail in \citet{Moustakas10} and \citet{Moustakas11}.  Briefly, we model the stellar
continuum as a non-negative linear combination of simple population
synthesis models of various ages.  We mask out the spectra at the
expected location of all emission lines and sky features and fit in an
iterative process that optimizes for the object's redshift and
velocity dispersion.  As part of this process, the residuals to this
fit - which contain continuum residuals and emission lines - are fit
to obtain emission line fluxes.  These emission line fits are then
subtracted from the observed spectra.  We individually inspected every spectrum to verify the quality of the fit and the decomposition of the absorption and emission-lines.  Some of the fits were able to be improved by flagging parts of the spectrum with sky residuals.  Nonetheless, we were forced to remove some galaxies with poor fits, most of which resulted from significant sky residuals being coincident with lines of interest. The result of this process are a set of
emission line measurements and a continuum spectrum that has been
``cleaned'' of emission.  We have multiple Balmer lines in our spectra
to independently constrain the absorption and emission and our
decomposition is therefore robust.

From these data products, we measure the equivalent width
of the \oii, \oiii, and Balmer emission lines.  These lines are used to
measure the SFR and to diagnose the presence of AGN activity.  We
also measure \ewhda\ and \ewhga\ as these lines are measures of the luminosity-weighted mean age and are useful
indicators of the presence of young and intermediate age stellar
populations.  Specifically we use the Lick continuum and line windows as defined in \citet{Worthey97} and \citet{Trager98}
to measure the Balmer absorption EWs.  We quantify the amount of Balmer absorption using the average of the \ewhda\ and \ewhga\  and call this index \hdga.  All index measurements are given in Table~\ref{indices_tab}.

\section{Galaxy selection}
\label{Sec:sel}

\subsection{Stellar Mass Completeness}
\label{Sec:mass_compl}

Galaxy properties depend strongly on stellar mass
\citep[e.g.][]{Kauffmann03} and environment
\citep[e.g.][]{Kauffmann04} and we must therefore control for the
former in order to study trends in the latter.  We do this by
selecting galaxies above a stellar mass limit for which we are
complete to galaxies of all stellar mass-to-light ratios \mlstar.  We
determine an empirical stellar mass limit following
\citet{Marchesini09} and \citet{Moustakas13}.  We take galaxies between 0.1 and 1.5 magnitudes
brighter than our observed magnitude limit at each redshift and scale
the stellar masses to the observed magnitude limit.  The upper mass
limit that encompasses 95\% of the scaled masses gives us an
indication of the maximum mass for which we are complete.  Essentially
this takes the maximum stellar mass-to-light ratio from our data and
assumes that it is indicative of that at the magnitude limit.  As
described in \citet{Marchesini09}, this technique has advantages over
traditional methods that use maximally old and unreddened stellar
populations \citep[e.g.][]{Dickinson03} in that it makes fewer
assumptions and derives the maximum \mlstar\ directly from the data.

The EDisCS spectroscopic magnitude limit is $I=22$ for $0.4<z<0.6$
clusters and $I=23$ for $0.6<z<0.8$ clusters.  The fainter selection
at high redshift offsets the increasing luminosity distance and the
same mass limit log(\mstar/\msol$)>10.4$, applies for all of our systems.
In all of what follows we only consider galaxies above this limit.  This results in a total of 163, 55, and 251 field, group and cluster galaxies respectively. In the last column of Table~\ref{clustdat_tab} we indicate the number of galaxies in this final sample that come from each cluster and group.

\begin{figure*}
\plotone{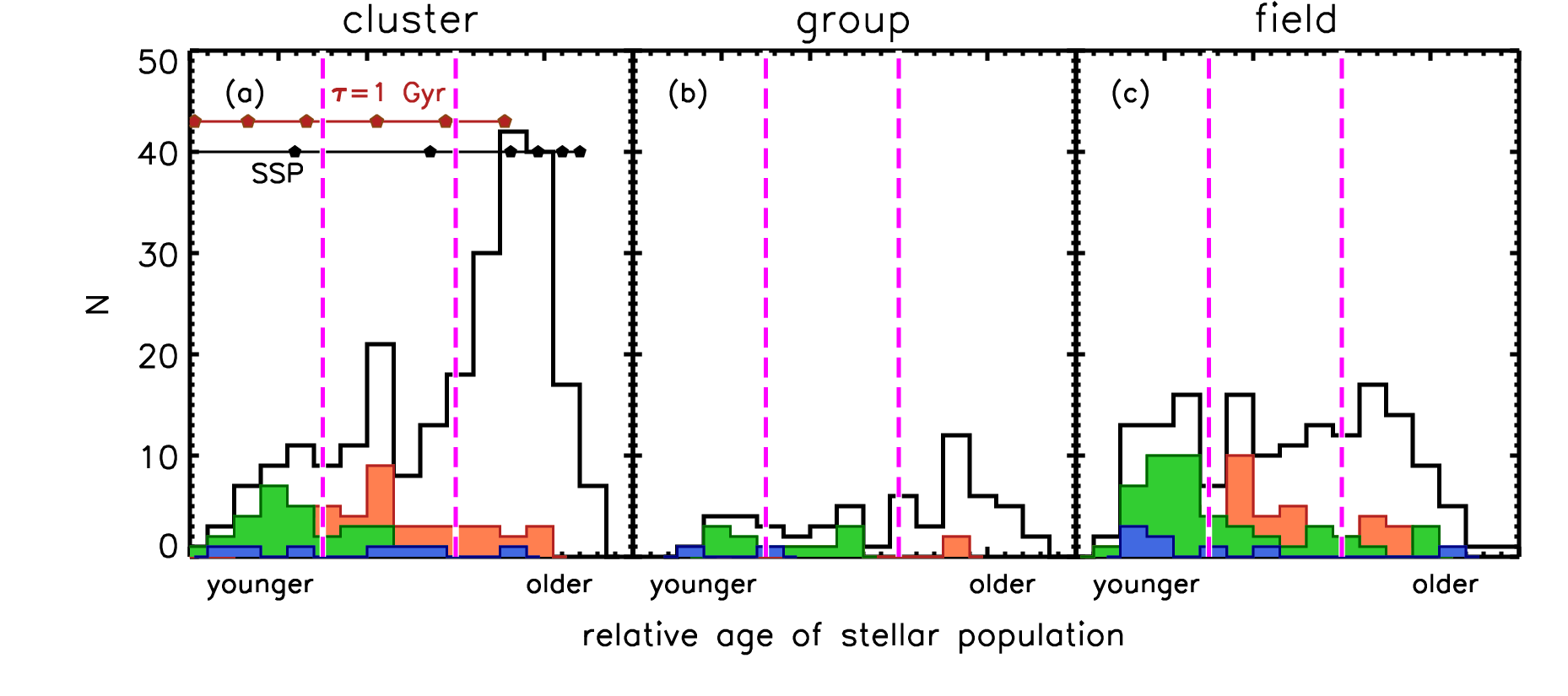}
\caption {A histogram of galaxies along the direction of the sequence
  in \hda\ and \hga\ absorption line strength (\hdga) and \dn\ (Fig.~\ref{Fig:d4hd}.)  The x-axis corresponds to the shortest distance of each object in Fig.~\ref{Fig:d4hd} to a zeropoint line in the far upper left of the panels in that plot that has an identical slope to the magenta lines shown in that figure.   The units are arbitrary but identical in each plot.  The different
  panels correspond to cluster, group, and field galaxies.  The
  vertical lines mark the bins in relative stellar population age that
  we label ``older'' (right bin), ``intermediate'' (middle bin), and
  ``younger'' (left bin). They are identical to the magenta lines shown in Fig.~\ref{Fig:d4hd}.  The colored histograms in each panel show the distribution of  galaxies with \ewoii $>$5\ang\ using the same color coding by emission line strength as in Fig.~\ref{Fig:d4hd}. The horizontal lines in panel (a) correspond to the single burst (black) and exponential model with $\tau=1Gyr$ (red) shown in Fig.~\ref{Fig:d4hd}.  The points on those lines mark 1~Gyr intervals increasing to the right.}
\label{Fig:d4hd_hist}
\end{figure*}

\begin{figure}
\epsscale{1.25}
\vspace{-1in}
\plotone{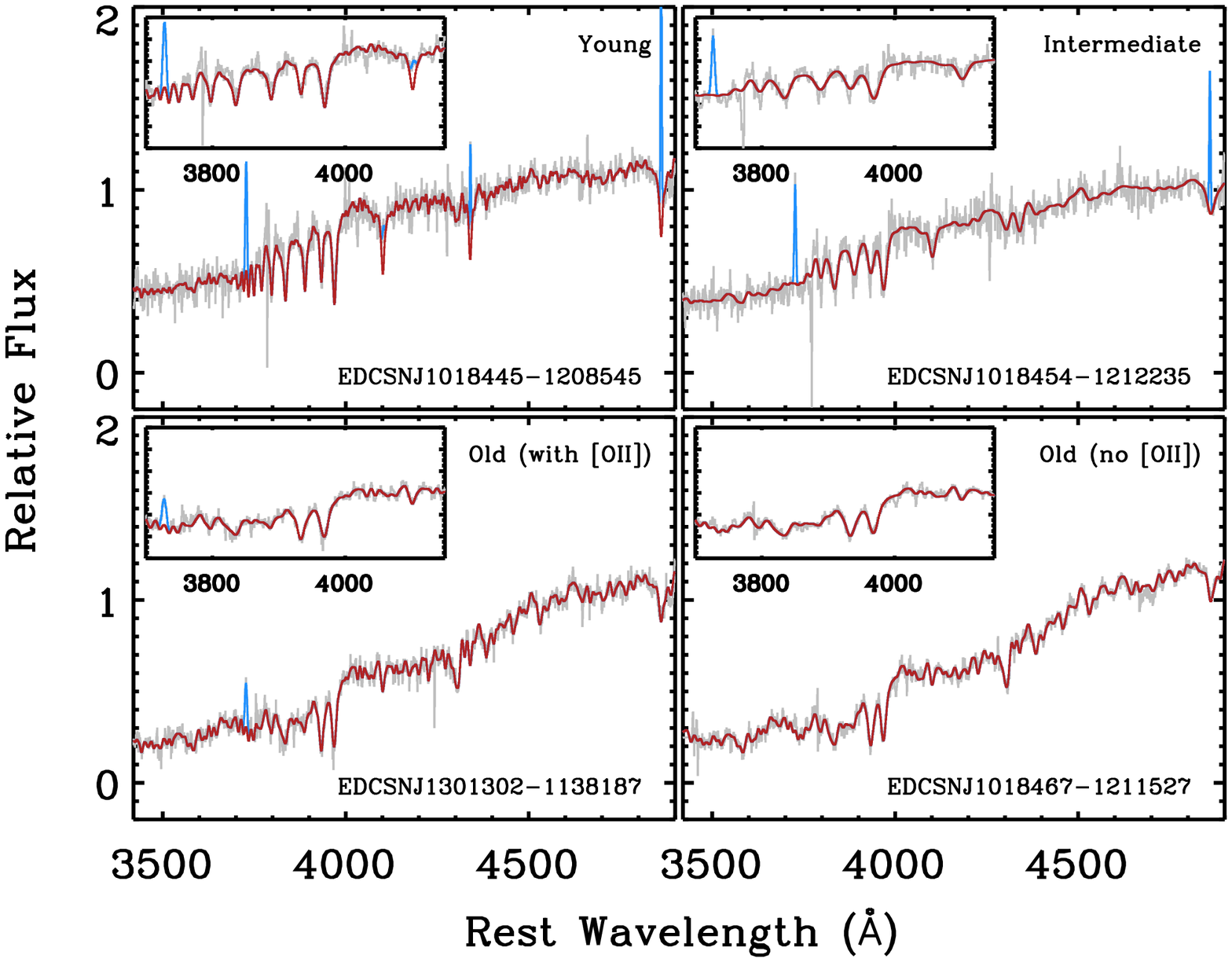}
\vspace{-1in}
\caption{ Typical example spectra for galaxies in each of our three different absorption line spectral classifications.  For the ``older" sample, we also show an example of a galaxy with an old stellar spectrum and with [\ion{O}{2}] emission.  The gray curve represents the data.  The red curve represents the model fit to the spectrum excluding the wavelengths were emission lines are expected.  Blue curves show significantly detected emission lines.  The inset in each panel is a zoom into the region around the H$\delta$ and H$\gamma$ lines and \dn.} 
\label{Fig:spec_examp}
\end{figure}

\subsection{Isolating Galaxies Dominated by Old Stars}
\label{Sec:agesel}

In Figure~\ref{Fig:d4hd} we show the \hdga\ strength
vs. \dn\ for  mass-selected cluster, group, and field
galaxies at $0.4<z<0.8$ with the points coded by \ewoii.  It is clear
that the galaxies in all environments follow a locus in this plot.  As
has been pointed out by many authors
\citep[e.g.][]{Rudnick00,Kauffmann03}, this is a sequence of changing
luminosity-weighted age in a galaxy stellar populations.  Galaxies
whose light is dominated by old stars appear in the lower right and
the fraction of young stars increasing to the upper left.  We
demonstrate this using model tracks in panel (d) of the figure,
derived from the \citet{Bruzual03} stellar population synthesis code.
These illustrative tracks represent a single stellar population formed
in a burst (SSP), a constant star formation history model, and
three exponentially declining SFHs with a timescale of 0.3, 1, and 2~Gyr, all
with solar metallicity.  All continuous SFHs lie in the same space as
the exponential tracks roughly independent of metallicity.  The models
clearly show that many e-foldings of the SFH are required to move a
galaxy into the lower right part of this diagram.  In practice this
requires that there are very few stars with intermediate or young ages
contributing to the galaxy light. 

Motivated by these results we attempt to draw boundaries in this
diagram which can separate the galaxies based on a relative measure of
the luminosity-weighted stellar age.  In Figure~\ref{Fig:d4hd_hist} we
show a histogram of the log(\mstar/\msol$)>10.4$ galaxies along this
sequence, running from the upper left to lower right.  The population
of older galaxies is clearly identifiable in the clusters and groups
as being the dominant population by numbers.  By contrast, in the
field the relative numbers of galaxies are roughly constant as a
function of age.  To isolate galaxies by the luminosity weighted age
of their stellar populations we delineate three regions along this
sequence.  The slope of these lines are 7.2.  We select galaxies whose light is dominated by old stars by
drawing a line in this histogram (and a corresponding diagonal line in
Figure~\ref{Fig:d4hd}) that separates the peak of old galaxies from
the rest.  Using the models from Figure~\ref{Fig:d4hd}d, we see that
this division (lower diagonal dashed line) selects galaxies that must
have little contributions from young or intermediate age stellar
populations, corresponding to $\sim 6$ e-foldings of the SFH.  In Figure~\ref{Fig:spec_examp} we show representative  spectra for our different classes and show one example each from the ``older" class of galaxies with and without emission.  The absorption spectra of the two ``older" classes show spectra dominated by old stars. 

\begin{figure}
\epsscale{1.1}
\plotone{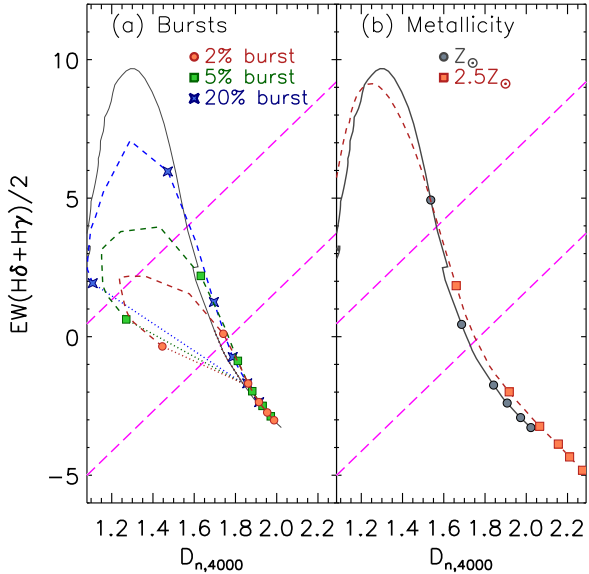}
\caption {\textit{Left panel --} the effect of a $10^8$year burst
  comprising 2, 5, and 20\% of the galaxy's stellar mass superimposed
  on a passively evolving stellar population.  The burst happens 3~Gyr
  after the initial epoch of star formation.  The symbols are spaced
  0.5~Gyr apart.  The dotted line shows the evolution at post-burst
  times less than 0.5~Gyr and the dashed lines show the later
  evolution.  For burst fractions in excess of 2\%, we will classify a
  galaxy as ``old'' if the star formation ceased at least 1~Gyr in the
  past.  \textit{Right panel --} the effect of metallicity on the
  indices for a passively evolving population.  The symbols are spaced
  1~Gyr apart.  Even at super-solar metallicities galaxies we will only
  classify a galaxy as ``old'' if we are seeing the star formation at
  least 1.5~Gyr after the cessation of star formation.}
\label{Fig:models}
\end{figure}

In Figure~\ref{Fig:models} we explore the effects of
bursts and metallicity on the location of objects in the \dn--\hdga\
 plane.  In order to meet our \dn--\hdga\ criteria for older
ages, no more than 2\% of the stars can have been formed in the past 1~Gyr from a burst superimposed on a 3~Gyr old passively evolving solar metallicity population.   We also explore the effect of metallicity as that will shift the models to larger \dn\ and smaller \hdga.  Even SSPs with a metallicity of 2.5$Z_\odot$ must have experienced their last major epoch of star formation at least 1.5~Gyr before the epoch of observation.  In comparison, SSPs with solar metallicity enter the lower-right region of the plot 2.5~Gyr after their formation.  Galaxies with a superimposed burst on an older population require less time to enter this region (Figure~\ref{Fig:models} left panel) because their is a substantial contribution to the total light from the pre-existing population, which accelerates the evolution in the \dn--\hdga\ plane.  As the
galaxies in this region are not truly old but merely have a lack of
young and intermediate age stellar populations we designate this the
``older'' region.

The remaining space is split between the ``younger'' and
``intermediate'' regions.  The dividing line between ``younger'' and
``intermediate'' has been chosen to be approximately between the \hdga\ and \dn\ values for a  6~Gyr old
population with a constant star formation rate model and those for an exponentially declining SFR model
with an exponential timescale of 2~Gyr.  Changing the exact location or slopes of the dividing lines does not significantly affect our results.

The age interpretation of our stellar index measurements is consistent
with the measured strength of \oii\ in our galaxies as the fraction of
\oii emitters and the characteristic \ewoii\ decreases towards older
stellar ages.  However, there is a population of emission line
galaxies at old ages.  They are the main focus of this paper and will
be discussed in subsequent sections.

\subsubsection{The Colors of Galaxies with Different Spectral Types}
\label{Sec:col}

\begin{figure*}[t]
\epsscale{0.9}
\plotone{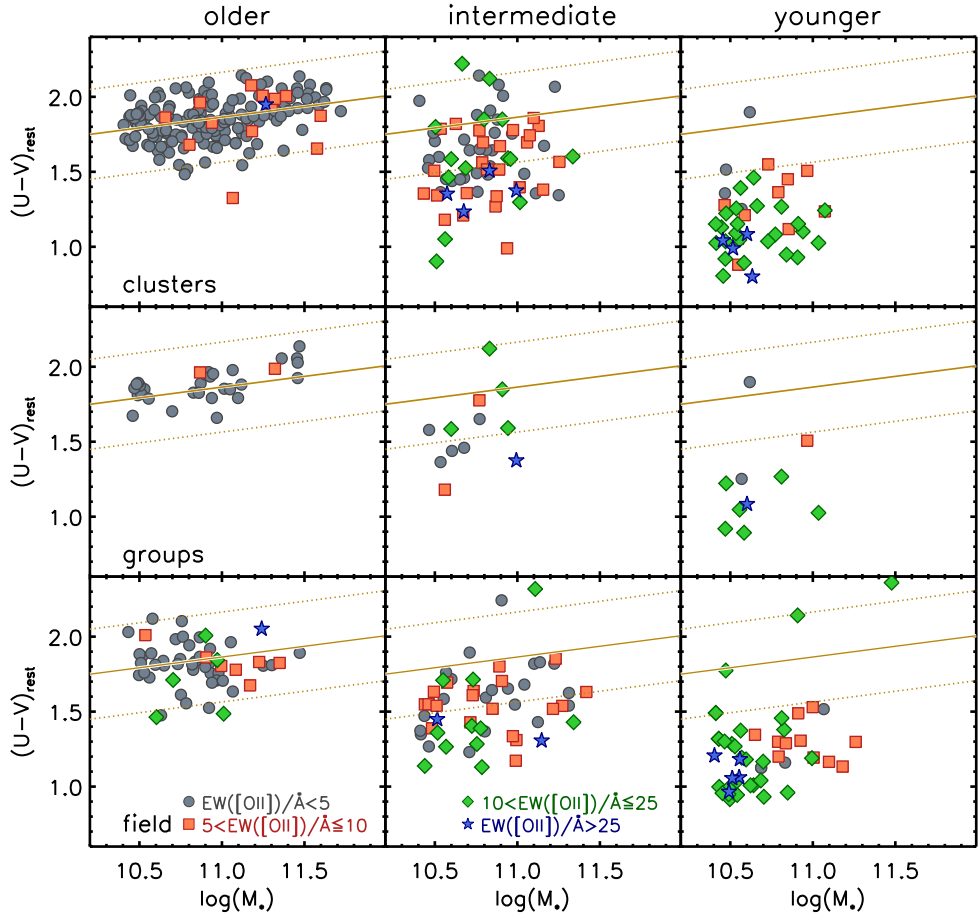}
\caption {A color-mass plot for cluster, group, and field galaxies
  divided into bins of relative age as described in the text.  The
  yellow solid lines are identical in each panel and show a robust
  linear fit to the ``older'' galaxies in clusters and the dotted
  lines are spaced at $\pm 0.3$ mag.  This encompasses nearly all of
  our ``older'' objects.  There are many objects on the red sequence
  with significant populations of intermediate age stars (middle
  column).  These galaxies are biased towards the blue side of the red
  sequence.  This implies that color selection in all environments
  probes a significant range in ages.  The ``older'' galaxies with
  emission - the focus of this paper - trace the whole sample of
  ``older'' galaxies.}
\label{Fig:colmass}
\end{figure*}

A major advantage of our indices is that they are less susceptible 
to dust than broad-band colors as they each probe a small wavelength baseline.  We
quantify this in Figure~\ref{Fig:colmass}, where we plot
color-magnitude diagrams for galaxies classified by environment,
\ewoii, and relative age of stellar populations (see
\S\ref{Sec:agesel}).  We define the red sequence by fitting to the
population of cluster galaxies with old spectroscopic ages and no
emission.  For all three environments these galaxies fall on the same red
sequence with equal scatter around this line.  For our
``intermediate'' age bin we find that many of the galaxies would be
classified as red sequence objects even though they have had
significant star formation within the past 1~Gyr and often have
emission lines.  Many of these red ``intermediate'' age galaxies have
significant obscuration, which drive their red colors (Rudnick, G. et
al. in prep).  Some subset may also be recently quenched.  It seems therefore clear that our spectroscopic selection,
when compared to a simple \textit{single} color selection, provides a
more robust way to select truly passive galaxies with no young stellar
populations.

\section{The Environmental Dependence of Galaxies with Old Stars and Weak Emission.}
\label{Sec:env_dep}

\begin{figure}
\epsscale{1.1}
\plotone{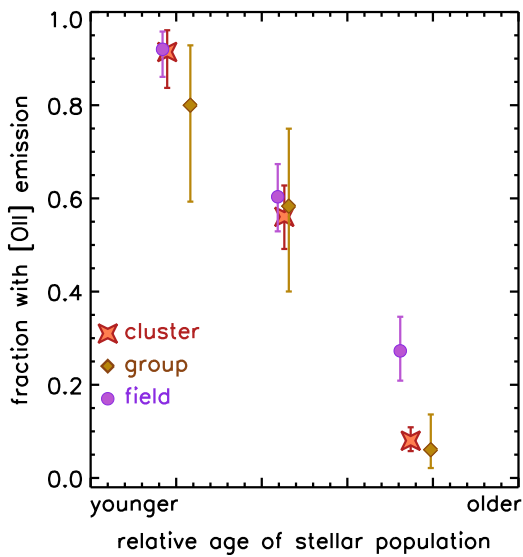}
\caption {The fraction of galaxies with \ewoii$>5$\ang\ as a function
  of relative stellar population age as defined in
  Fig.~\ref{Fig:d4hd} and Fig.~\ref{Fig:d4hd_hist}. The points have been shifted in the x-direction slightly with respect to one another so that they do not overlap.  This plot demonstrates one of the key results of the paper, namely that old galaxies in the field have a higher fraction of \oii\ emission than old galaxies in clusters and groups.}
\label{Fig:emfrac}
\end{figure}

In Figures~\ref{Fig:d4hd}, \ref{Fig:d4hd_hist}, and \ref{Fig:colmass},
it is apparent that there are galaxies with weak emission lines and
old stellar ages.  These emission-line galaxies are present in roughly
equal numbers in the field and in clusters and are nearly absent in groups.
Already from these figures you can see that the very different total
number of ``older'' galaxies implies that the fraction of such
galaxies with emission must depend greatly on environment.  We
quantify this result in Figure~\ref{Fig:emfrac} where we plot the fraction of
galaxies with \oii\ emission as a function of age.  Among the three
environments, there are no differences in the fractions for galaxies
with ``younger'' or ``intermediate'' ages.  However, for the ``older''
population of galaxies, there is a difference in the fraction with
\ewoii$>5$\ang, which we call $f_{oe}$.  5\ang is the equivalent width limit above which we can reliably identify and fit the \oii\ line.

In the field the fraction of ``older" galaxies with emission is
$f_{oe,field}=0.27^{+0.07}_{-0.06}$ while in clusters and groups
$f_{oe,clusters}=0.08^{+0.03}_{-0.02}$ and $f_{oe,groups}=0.06^{+0.08}_{-0.04}$ respectively.
Phrased differently, the field has a 2.7$\sigma$ higher fraction of
old emission line galaxies when compared to the cluster galaxies and a
2.1$\sigma$ higher fraction when compared to group galaxies.  When combining group and cluster galaxies together, the difference with respect to the field changes to a 2.8$\sigma$ difference.  We provide the fractions of \oii\ emitters in Table~\ref{stellpops_tab}.  It is important to remember that our field galaxies are excluded from lying in groups or clusters and therefore indicate a true field sample.

\setcounter{table}{2}
\ifemulate
	\begin{deluxetable*}{clllllll}
\else
	\begin{deluxetable}{clllllll}
\fi
\tablecaption{Fractions of emission-line galaxies as a function of stellar age and environment}
\tablewidth{0pt}
\tablehead{\colhead{Relative Age\tablenotemark{a}} & \colhead{$f_{e,field}$\tablenotemark{b}} &  \colhead{$f_{e,group}$\tablenotemark{b}}  & \colhead{$f_{e,cluster}$\tablenotemark{b}} &  \colhead{$f_{e,group+cluster}$}\tablenotemark{b} & $N_{field}$ & $N_{group}$ & $N_{cluster}$}
\startdata
younger &  $0.92_{-0.06}^{+0.04}$ & $0.80_{-0.21}^{+0.13}$ & $0.91_{-0.08}^{+0.05}$ & $0.89_{-0.07}^{+0.05}$ & 50 & 10 & 35\\\\
intermediate & $0.60_{-0.07}^{+0.07}$ & $0.58_{-0.18}^{+0.17}$ & $0.56_{-0.07}^{+0.07}$ & $0.56_{-0.06}^{+0.06}$ & 58 & 12 & 66\\\\
older & $0.27_{-0.06}^{+0.07}$ & $0.06_{-0.04}^{+0.08}$ & $0.08_{-0.02}^{+0.03}$ & $0.08_{-0.02}^{+0.02}$ & 55 & 33 & 150
 \enddata
\label{stellpops_tab}
\tablenotetext{a}{The relative ages corresponding to the age divisions in \dn-\hdga\ plane as shown in Figure~\ref{Fig:d4hd} and \ref{Fig:d4hd_hist}. }
\tablenotetext{b}{The fraction of galaxies in each relative age bin and in each environment that have \ewoii$>5$\ang.}
\tablecomments{All errors on the fractions are computed using binomial errors.  Galaxies are limited in stellar mass to  log(\mstar/\msol$)>10.4.$.  The numbers in the rightmost columns correspond to the total number of galaxies in each age-environment combination that pass all of our selection criteria.}
\ifemulate
	\end{deluxetable*}
\else
	\end{deluxetable}
\fi

We test for a stellar mass dependence in the emission-line fractions.
We divide the sample into two bins of stellar mass split at
log(\mstar/\msol$)=11.0$ and we find that the results in the two stellar
mass bins are consistent with results for the whole sample, although
with lower statistical significance.  This conclusion does not depend
on the exact definition of the stellar mass bins.  We have also tested to see if the stellar mass distributions of the samples in different environments are different.  Using a K-S test, the stellar mass distributions for all ``older" galaxies in clusters and the field have a 13\% probability of being drawn from the same underlying distribution.  Likewise, the ``older" emission-line galaxies in clusters and the field have a 18\% probability of being drawn from the same distribution. 

\begin{figure}
\epsscale{1.1}
\plotone{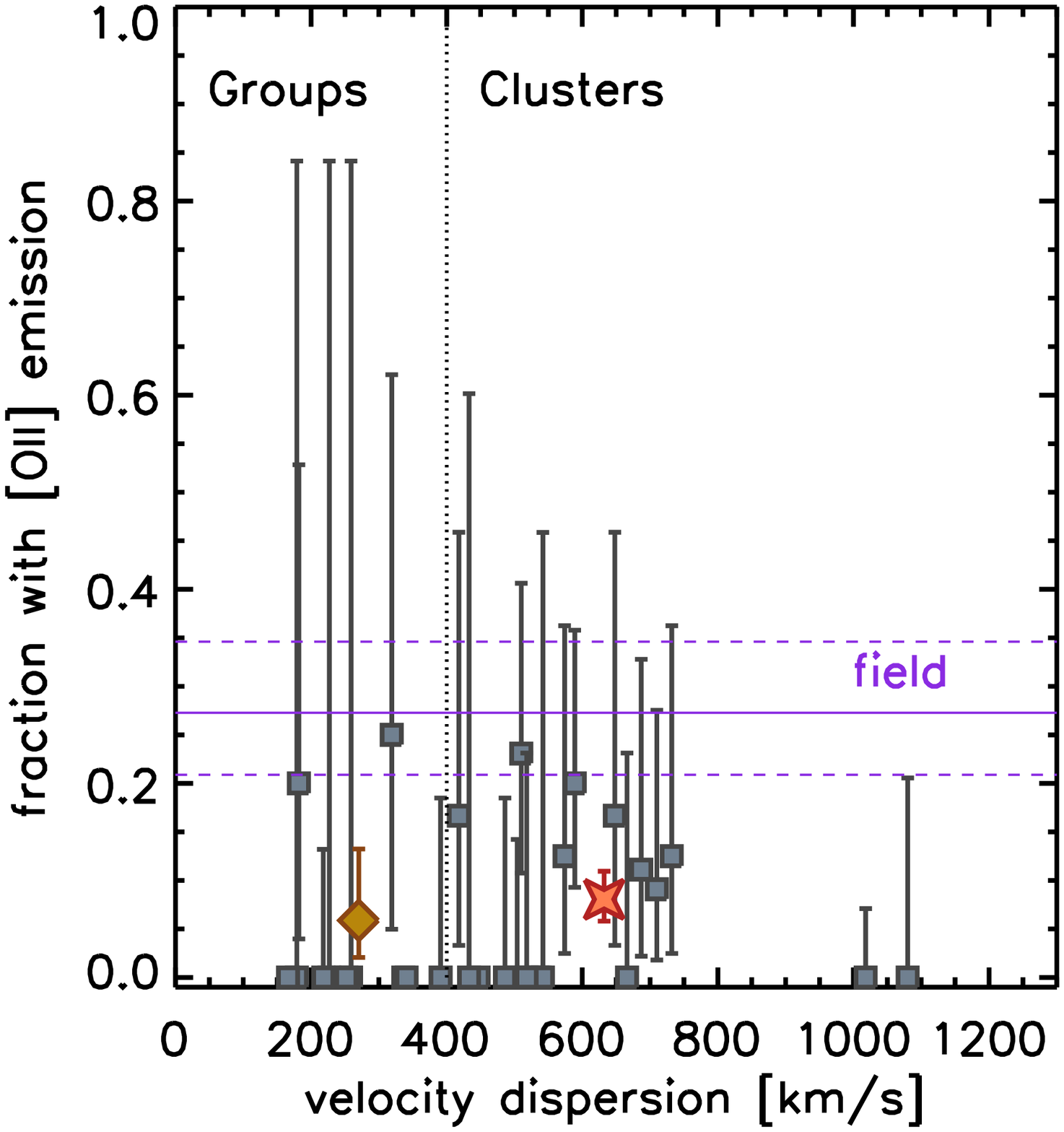}
\caption {The fraction of ``older'' galaxies with \oii\ emission for
  all of our groups and clusters and for the field.  The gray squares
  are for the individual systems with groups and clusters being divided
  at 400~km/s.  The large yellow diamond and large red star are the
  values for all the galaxies in the group and cluster samples respectively.
  The horizontal purple lines indicates the value for the field and its
  68\% confidence interval.}
\label{Fig:emfrac_sig}
\end{figure}

A potential concern is that our result may be dominated by a subset of
the systems in each environmental bin.  In Figure~\ref{Fig:emfrac_sig}
we address this by showing the emission line fractions among ``older''
galaxies for each system independently.  We find that all of the
individual clusters and groups are below the value of the field
implying that our result is true for the whole population of galaxies
and is not just dominated by a few specific structures.  It is perhaps
interesting that the two most massive clusters show a lower fraction
of \oii\ emitters than lower mass clusters.  This is reminiscent of
the results of \citet{Poggianti06}, who used the same cluster sample
as in this work to conclude that the fraction of \oii\ emitters for
galaxies of all ages were lower for more massive clusters.  Although
their measures of \ewoii\ are systematically different from ours and
were conducted on a luminosity-selected sample, it is still
interesting that this mass dependence may persist even for galaxies
such as ours with extremely weak emission and old stellar populations.

\begin{figure}
\epsscale{1.10}
\plotone{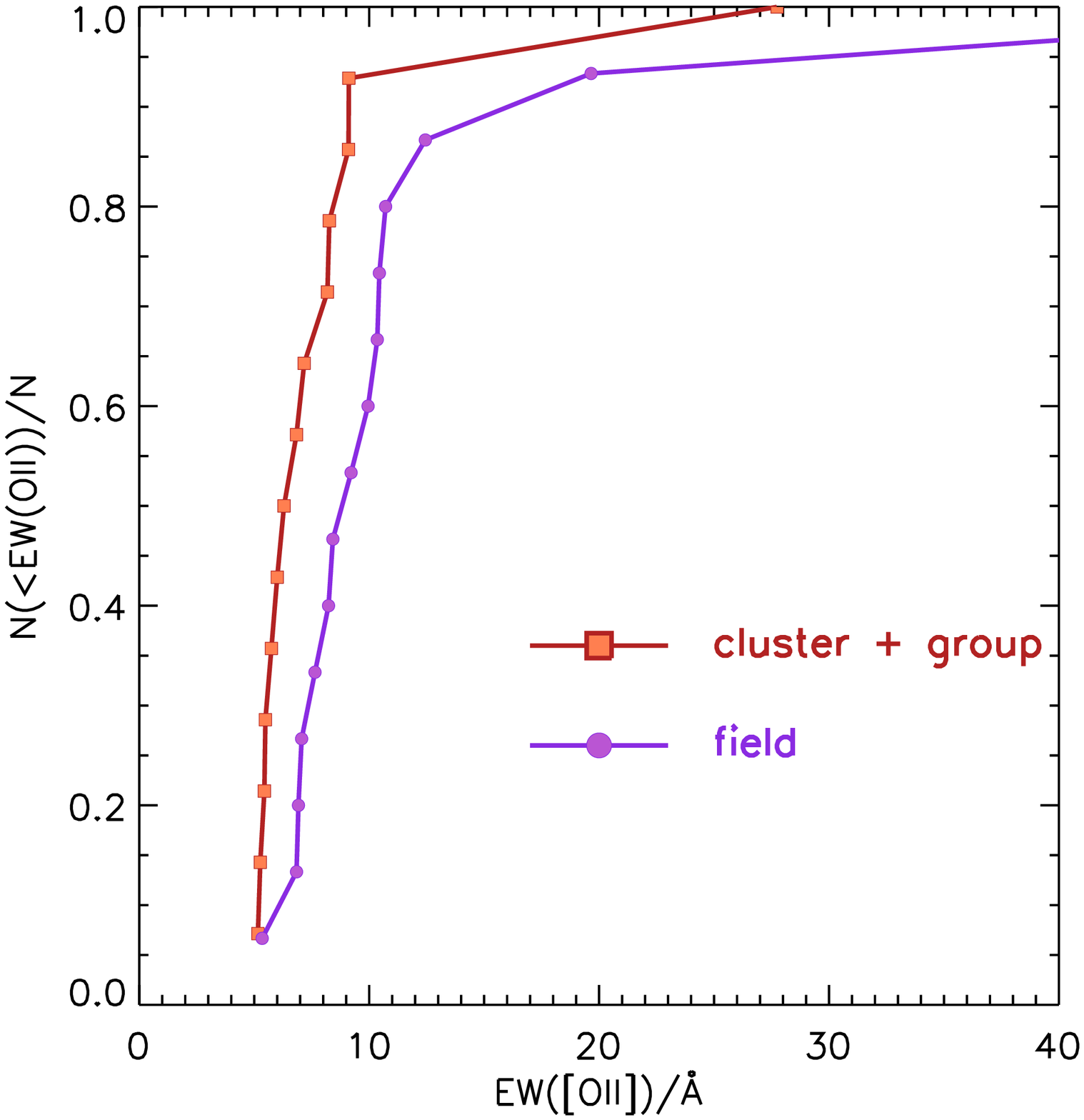}
\caption {The cumulative distributions of \ewoii\ for galaxies with
  \ewoii$>5$\ang\ that are classified as ``older'' based on \dn\ and
  the \hdga\ strength.    The cluster$+$group distribution is shifted towards lower
  \ewoii\ values, with only a 3\% K-S probability of being drawn from
  the same distribution as the field galaxies.}
\label{Fig:em_cumhist}
\end{figure}

In addition to measuring the fraction of emission it is also
instructive to look at the distribution of \ewoii.  In
Figure~\ref{Fig:em_cumhist} we show the distribution of \ewoii\ in our
mass-selected sample of ``older'' galaxies with \ewoii$>5$\ang.  As
there are almost no ``older'' galaxies in our groups that meet this
\ewoii\ threshold, we can only examine the distribution of clusters
and the field.  The cluster galaxies are shifted to systematically
lower \ewoii\ values than the field and the K-S probability that they
are drawn from the same distribution is 3\%.  The median \ewoii\ of
the cluster and field population are slightly different at 6.3 and
8.8\ang\ respectively but the 68\% limits on the two distributions
overlap significantly.  It is therefore difficult to draw any strong
conclusions about the distribution of \ewoii\ in the different
environments but we will discuss a possible origin for this shift in
\S\ref{Sec:disc}.

\subsection{Radial and velocity distributions of different populations}
\label{Sec:phase}

\begin{figure}
\epsscale{1.20}
\plotone{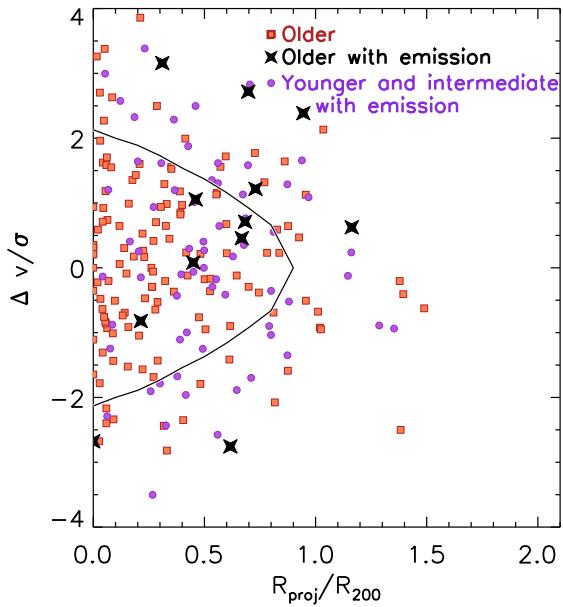}
\caption {The projected phase-space positions
  of different cluster galaxy populations.  The solid curve is from
  \citet{Mahajan11} and contains most of the virialized galaxies in
  their simulations. The purple points refer to those galaxies in the ``younger" and ``intermediate" age categories.  }
\label{Fig:rad_v}
\end{figure}

\begin{figure*}
\epsscale{1.10}
\plottwo{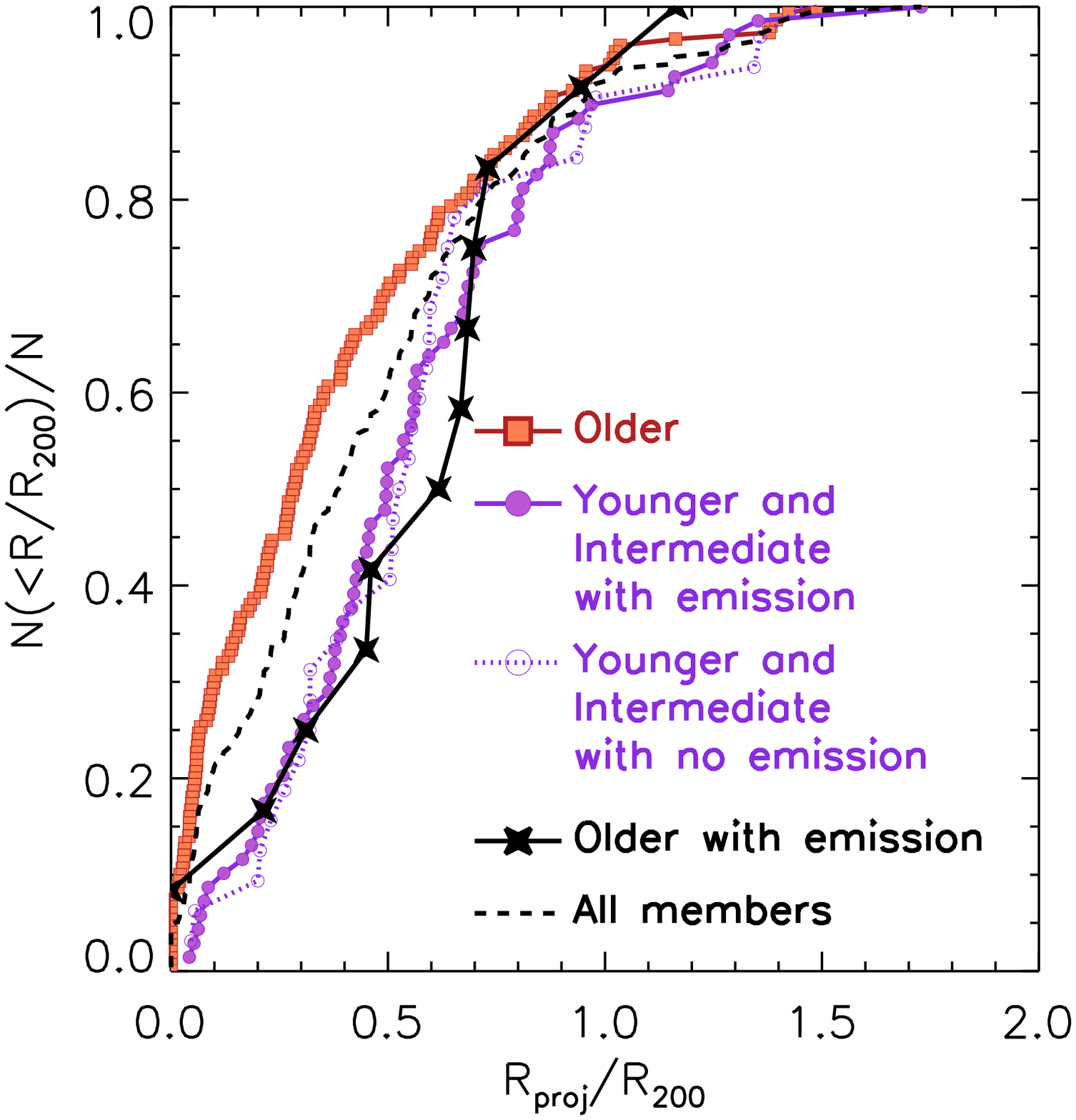}{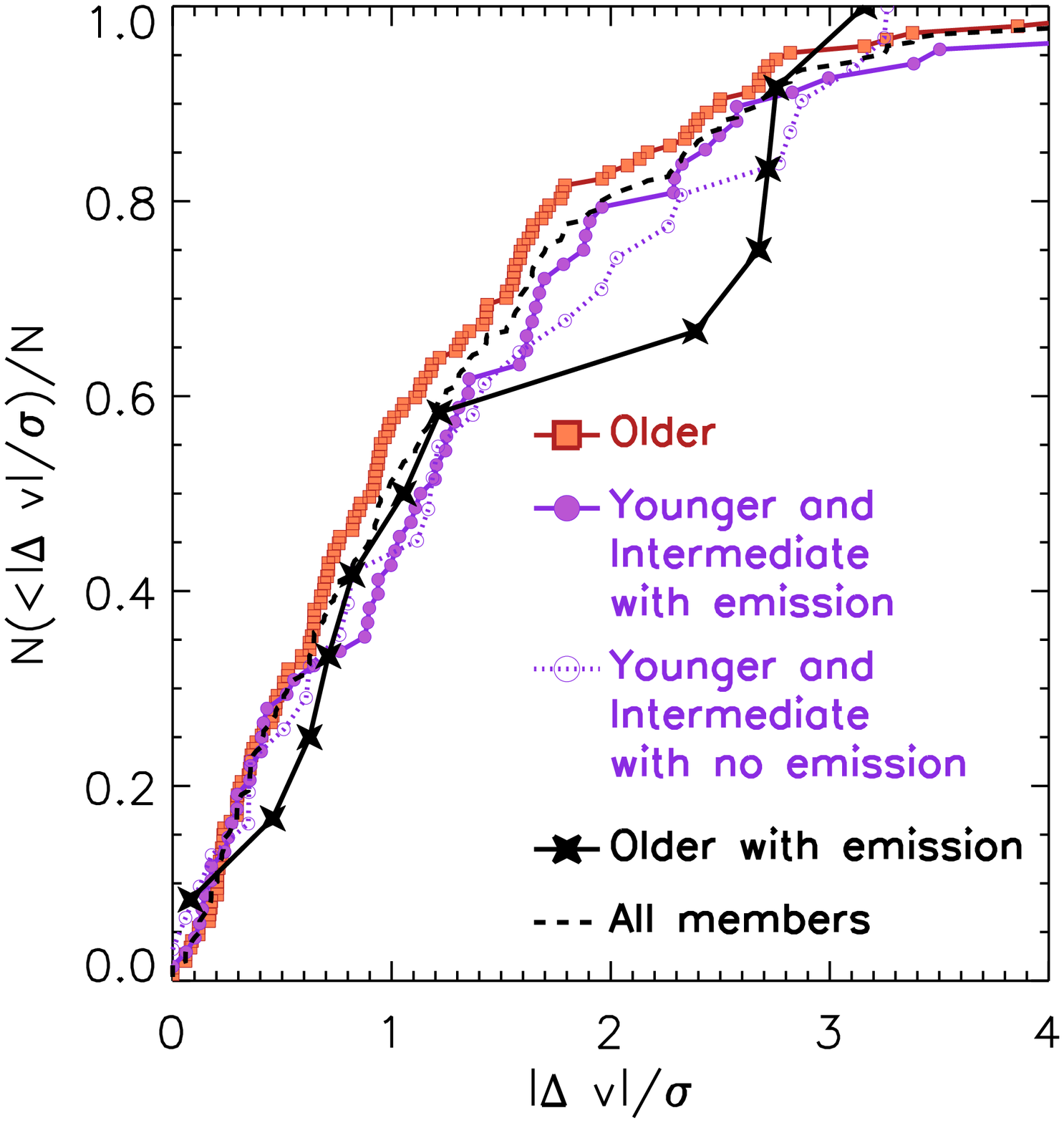}
\caption {\textit{Left panel --} The
  cumulative histogram of $R_{proj}/R_{200}$
  for the different galaxy populations.  All galaxies with emission
  are systematically biased  to large cluster-centric radii.  \textit{Right panel --} The cumulative histogram of $|\Delta v| / \sigma$ for the same galaxy populations.  All of the populations are statistically consistent with being drawn from the same underlying distribution.}
\label{Fig:rad_v_cumhist}
\end{figure*}

In order to gain insights as to the infall history of the different
populations in Figure~\ref{Fig:rad_v} we examine the projected
phase-space locations of the different galaxy populations.  We limit ourselves to clusters as they are the only systems with enough members for us to reliably determine $\sigma$ and hence $R_{200}$.  We also remove the cluster cl1138.2-1133a as it is too off-center in our spectroscopic observations to probe out to $R_{200}$.  This
diagram has been used by multiple authors to understand the
orbital state of cluster galaxies
\citep[e.g.][]{Mahajan11,Haines12,Oman13}.  In those papers the
authors used simulations to establish that the regions with small
projected radii ($R_{proj}/R_{200}<1.0$) and small velocities
($|\Delta v|/\sigma<2$) contain most of the virialized galaxies while
moving towards higher velocities or larger radii correspond to
galaxies that are either infalling for the first time or have just
finished their first pass through the cluster center (the latter are often called ``back-splash'' galaxies.)  In Figure~\ref{Fig:rad_v} we indicate the
corresponding virialized region from \citet{Mahajan11}.  One must exercise caution in tying locations in this diagram directly to a time since infall as \citet{Oman13} showed that the distribution of infall times at any given projected position-velocity space is very broad.  Nonetheless, galaxies lying within the curve in the left panel of Figure~\ref{Fig:rad_v} are strongly preferred to have been in the cluster for greater than 3~Gyr.  In figure~\ref{Fig:rad_v_cumhist}, we also plot the cumulative distributions of the different populations in both projected radius in units of $R_{200}$ and in radial velocity in units of the cluster velocity dispersion.

With the caveats above in mind, we find that
our ``older'' population preferentially occupies the virialized region of the
diagram.  In contrast, both the ``younger" and ``intermediate" galaxies, regardless of the presence of emission lines, and ``older''
emission line galaxies are largely absent from regions with low
$R_{proj}<R_{200}$ (Figure~\ref{Fig:rad_v_cumhist}).  
We test the significance of the distributions in phase space and find that the ``younger''$+$``intermediate"
and ``older'' galaxies have a $<2\times 10^{-3}$ K-S probability of having
been drawn from the same distribution in projected cluster-centric radius.  In contrast, the
``older'' emission-line galaxies are completely consistent with the
``younger''$+$``intermediate" galaxies and only marginally so ($P_{K-S}\approx3.1\%$) with the
``older'' galaxies.  A K-S test on the absolute value of the velocity relative to the
cluster velocity dispersion (Figure~\ref{Fig:rad_v_cumhist}; right panel) shows that all of the populations are consistent at the greater than 17\% level with being drawn from the same population.  We perform a third test in which we compute the shortest (or perpendicular) distance of each point to the nearest segment of the \citet{Mahajan11} curve, where this distance is measured in the axis units of Fig.~\ref{Fig:rad_v}, and assign negative distances to those galaxies inside the curve.  This is a measure in phase space of the distance from the virial region and combines velocity and position.  A K-S test of the distributions of this distance show that the ``older" emission-line galaxies and ``younger" galaxies are completely consistent while the ``older" emission line galaxies are only marginally consistent with general ``older" population ($P_{K-S}\approx6.7\%$).  As with the pure radial distributions, the ``younger'' and ``older'' galaxies have a $<2\times 10^{-3}$ K-S probability of being drawn from the same distribution.

\begin{figure*}
\plotone{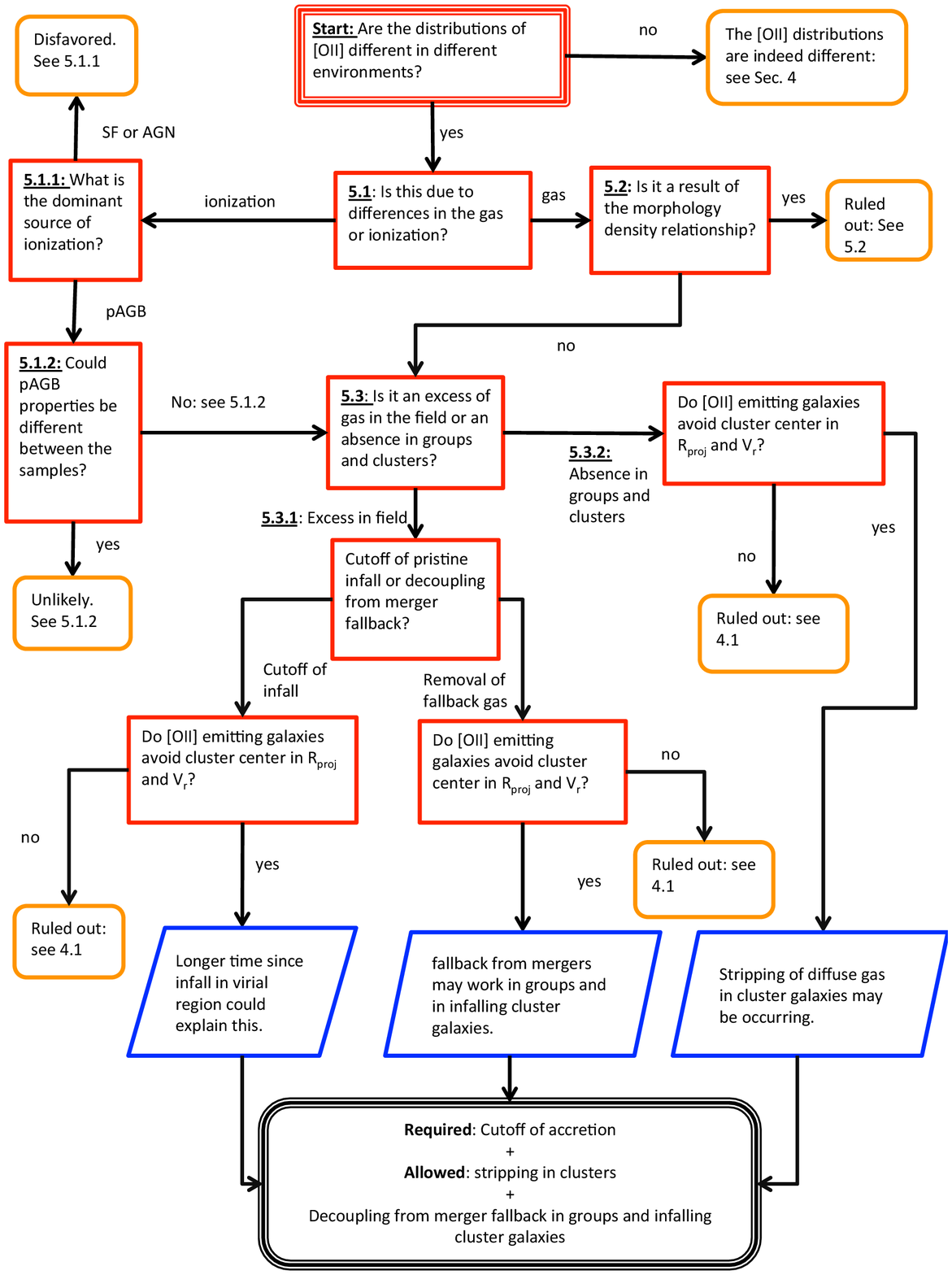}
\caption{This demonstrates the decision tree used in our discussion and outlines all of the major decisions made in coming to our conclusions.  The tree starts at the top middle of the box and the box representing our conclusion is at the lower right.  Red boxes represent decision branches, black arrows and accompanying text outline possibilities for each decision, yellow boxes represent that a given choice is not matched by observations, and blue box parallelograms provide information that goes with each choice.  \textbf{\underline{Bold and underlined}} text represent the section of the text that refers to each portion of the tree.}
\label{Fig:dec_tree}
\end{figure*}

\section{Discussion}
\label{Sec:disc}

As outlined in the previous sections, we have found that galaxies dominated by old stellar populations are more likely to have [\ion{O}{2}] emission in field environments than in group or clusters.  The first impulse is to attribute this effect to an environmental signature, given the differences with environment.  However, in determining the origin and cause of the observed difference in emission we examine a range of possibilities.  Our goal is not simply to pick one scenario that is consistent with our observed environmental dependence in the \ewoii\ distribution, but rather to test a large range of possible explanations and see what is required by the data, what is allowed, and what is ruled out.  Given the subtle nature of environmental effects, we feel it is important to carefully examine many alternative possibilities.  We proceed using a decision tree that outlines the path to our conclusions.  A graphical representation of this tree is given in Figure~\ref{Fig:dec_tree}, which identifies the sections that refer to each branch of the tree.  We now address these in turn starting from the top of the tree.

\subsection{Are the differences in [\ion{O}{2}] fraction driven by gas content or excitation?}

In this subsection we will first address the source of the excitation.  We will then discuss if this excitation source can be different in different environments and we will close by concluding that differences in gas content must be driving the differences in the [\ion{O}{2}] fractions.

\subsubsection{Source of the excitation}
\label{Sec:ion_source}

\begin{figure}
\epsscale{1.20}
\plotone{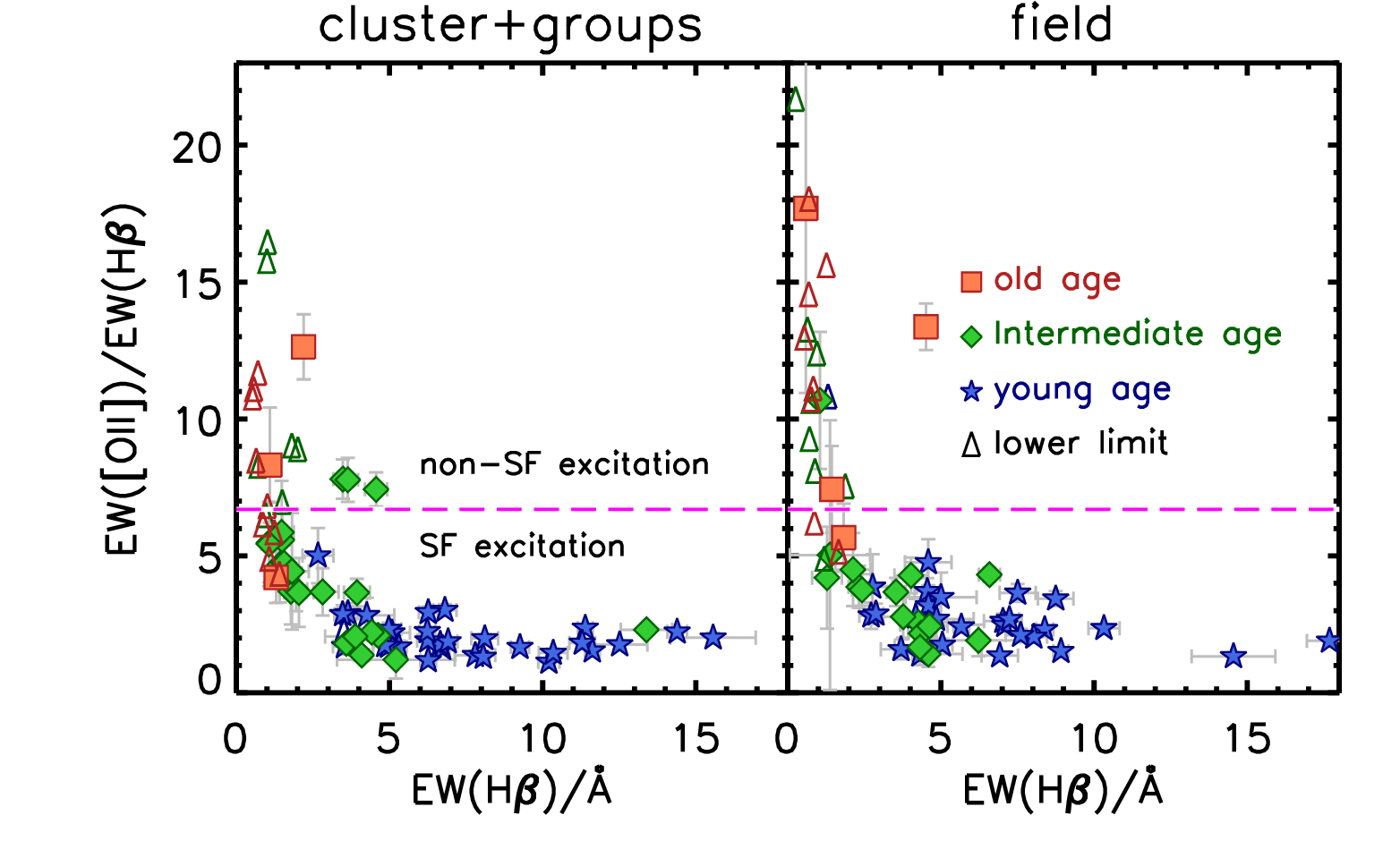}
\caption {An AGN diagnostic diagram for all of the galaxies in our
  sample with \ewoii$>5$\ang\ for which $H\beta$ could have been
  observed.  All emission and absorption features have been decomposed
  as described in the text.  Arrows are $1\sigma$ lower limits on the
  ratio, plotted at the $1\sigma$ upper limit of \ewhbe.  Line ratios
  above the horizontal line cannot be powered by normal star formation
  \citep{Sanchez09}.}
\label{Fig:agn_diag}
\end{figure}

Local studies have made significant progress towards understanding the
sources of gas excitation in red galaxies.  Many line diagnostics
have been used to study the emission mechanisms in red local galaxies
\citep[e.g.][]{Yan06,Sarzi06,Stasinska08,Sarzi10,Stasinska15} but as only $\approx 50\%$ our spectra with an \oii\ detection  go to long
enough wavelengths to measure even H$\beta$ we cannot use the same
line diagnostics as those authors for all our sources.  Instead we use a modified version
of the \citet{Yan06} diagnostic that was developed by
\citet{Sanchez09}.  This diagnostic works because star formation under
normal extinction conditions is not able to produce
\ewoii/\ewhbe$>6.7$ \citep[e.g.][]{Moustakas06a,Sanchez09}.  We plot this
ratio as a function of \ewhbe\ for our sources in
Figure~\ref{Fig:agn_diag}.  We include 1$\sigma$ lower limits on the
ratio for galaxies with no detected H$\beta$.  We see that the
``older'' galaxies are near to or above this line, implying that the
excitation is likely not caused by star formation.  We also verify that these results are unchanged if we plot the ratio of \oii\ to H$\beta$  fluxes instead of equivalent widths.

We further test the lack of star formation by searching for MIPS 24\micron\ detections in our ``older" galaxies with emission lines, using the deep MIPS data on our clusters from \citet{Finn10}.  These data have an infrared luminosity (\lir) 80\% completeness limit of $8.1\times 10^{10}$\lsol, corresponding to a SFR limit of $\approx 8~$\msol$/$year.  We find that only one of the ``older" cluster galaxies with \oii\ emission (8\%) and one in the field (7\%) have a MIPS 24\micron\ detection, while zero of our ``older" emission line galaxies are detected at 24\micron\ in groups.  In contrast, we find between 43-71\% of our ``intermediate" galaxies with \oii\ emission lines are detected with MIPS in the different environments, while 69-75\% of our ``younger" galaxies with \oii\ emission are detected, consistent with the ``intermediate" and ``younger" galaxies having significant ongoing star formation.  The \ewoii\ values for ``older" galaxies are lower than for the other two age bins.  We therefore also test whether this difference in the \ewoii\ values can account for the different fraction of MIPS detections.  We find that those galaxies with $5<$\ewoii$/$\ang$<10$ in the ``intermediate" and ``younger" age groups have MIPS detection fractions between 33-100\% depending on the age and environment bin.  This is significantly higher than for ``older" galaxies in any environment and indicates that weak \oii\ emitters in the ``intermediate" and ``younger" age groups, regardless of environment, are more likely to have their \oii\ emission powered by star formation.  Likewise, this implies that the \oii\ emission in ``older" galaxies is not coming from highly obscured star formation embedded in an older stellar cocoon. 

This is consistent with studies of the excitation source in local red
galaxies.  \citet{Yan06} found that many of the red galaxies in thee
SDSS with weak \oii\ emission (similar in strength to our \ewoii)
had line ratios consistent with a LINER-like spectrum \citep{Heckman80}.
\citet{Sarzi06} and \citet{Sarzi10} used SAURON integral field spectroscopy of nearby
early-type galaxies to demonstrate that much of the ionized emission
in early-types with a LINER-like spectrum is extended and traces the
stellar light.  They therefore concluded that the heating source is
consistent with post asymptotic giant branch  (pAGB) stars and not an AGN.\footnote{There are multiple potential sources of UV ionization and excitation in passive galaxies, including extremely blue horizontal branch stars, and multiple flavors of post-AGB stars \citep{Greggio90,Brown08}.  The relative contributions of these populations to the ionizing budget in early types is poorly constrained \citep{Brown08}.  For the purpose of this discussion we simplify by assuming pAGB stars perform the bulk of the excitation in our galaxies, although our conclusions are not dependent on the exact sub-population responsible for the ionization and excitation.  Note also that pAGB stars are to be distinguished from thermally pulsating AGB (TPAGB) stars, as the latter are dust enshrouded stars on the AGB while pAGB stars are the UV bright cores of evolved stars made visible via the expulsion of gas in a planetary nebula phase.}
  This finding is consistent with work by \citet{Annibali10}, who used long-slit spectroscopy of local early types, with \citet{Yan12}, who used SDSS spectroscopy of 500 passive galaxies observed as part of the Palomar Survey \citep{Ho95}, and \citet{Singh13} and \citet{Kehrig12}, who studied the spatially resolved emission-line properties of galaxies in the CALIFA survey \citep{Sanchez12}.  Shocks are not a likely excitation source for this emission as the necessary shock velocities are not consistent with the observed gas kinematics \citep{Sarzi10,Yan12}.  

\begin{figure}
\epsscale{1.20}
\plotone{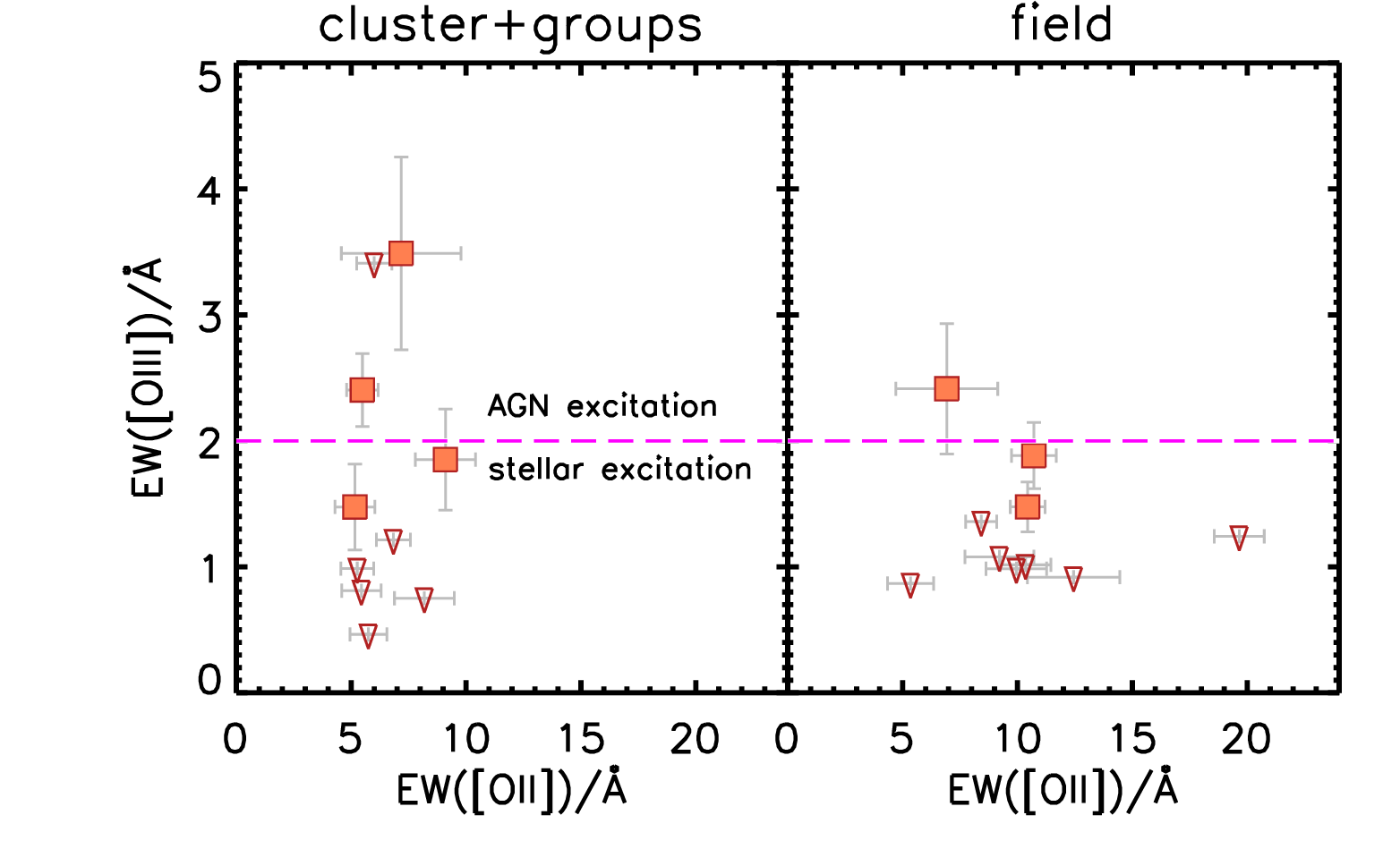}
\caption {A comparison of \ewoii\ and \ewoiii\ for all of the ``older" galaxies
  in our sample with \ewoii$>5$\ang\ for which \oiii\ could have been
  observed.  Points with arrows are $1\sigma$ upper limits on \ewoiii,
  plotted at the $1\sigma$ upper limit of \ewoiii.  \ewoiii\ below the
  horizontal line are consistent with pAGB heating while those above
  require an AGN.}
\label{Fig:o3_diag}
\end{figure}

To further test if our sources are consistent with a pAGB emission mechanism, we compare the \oiii\ strength from \citet{Sarzi10} with that predicted for our galaxies.  We do this as we only have access to \oiii\ for a handful of sources (but see below) and because \citet{Sarzi10} do not have access to \oii.  We start with the typical \ewoiii\ values of 1--2\AA\ from
\citet{Sarzi10}.  We then use the LINER line ratios from
\citet{Heckman80} who find that $L($\oii$)/L($\oiii$)\sim 2-4$.  Given
the typical continuum ratios between \oiii\ and \oii\ in our ``older''
galaxies of $\sim3$, this implies \ewoii\ for LINER like galaxies of
$6-24$\AA, which is consistent with the distribution of \ewoii\ seen
in Figure~\ref{Fig:em_cumhist}.  This implies that our emission is
consistent with a LINER-like spectrum that is produced by pAGB heating.  As
an additional check, in Figure~\ref{Fig:o3_diag} we show that
\ewoiii\ for our ``older'' galaxies has values typically consistent
with being powered by pAGB stars (\ewoiii$<2$\AA), with about $\sim 20\%$ of the
galaxies having \ewoiii\ that implies an AGN contribution.  If we instead compare $f($\oii) to $f($\oiii) we find that 30\% of our sources have \ewoiii\ that implies an AGN contribution.  The
spatial resolution of our spectroscopy is not high enough to address the
spatial extent of the emission and we can therefore only conclude that
the emission is likely not related to star formation but comes either
from a weak AGN or from the heating of gas by pAGB stars.

That the emission does not come from star formation in our ``older" galaxies is also consistent with the observed values of \dn\ and \hdga, as we show in Figure~\ref{Fig:models} that even a
small amount of ongoing star formation is enough to move our galaxies
out the ``older'' bin.  As a final support that the weak \oii\ emission is not coming from star formation, \citet{Sanchez09} used indices to measure the ages of red sequence galaxies in our same clusters - indeed for many of the same galaxies - and find that the luminosity-weighted mean stellar age of their cluster galaxies does not change when excluding those red sequence galaxies with weak emission.  This further bolsters our claim that the emission is not coming from small amounts of star formation.   It is of course possible that galaxies with
small amounts of star formation could temporarily leave our ``older''
bin and be excluded from our analysis but this does not alter the
significant difference in the \oii\ fraction among older galaxies seen in different environments.

\citet{Jaffe14} also analyzed morphologically selected early-type galaxies in EDisCS with extended emission.  They concluded that the emission from their sample was powered predominantly by star formation and not by evolved stellar populations.  This conclusion was based both on the blue colors of the galaxies and because they fell predominantly in the ``intermediate" and ``younger" section defined by this paper in the \dn-\hdga\ plane.  Indeed, only 4 of their galaxies would have entered our ``older" sample.  It is also worth noting that 10 of their 18 early-type galaxies have have \ewoiii$>10$\ang, and 4 would have not met our \ewoii$>5$\ang limit.  Therefore 4/14=30\% of their sample has 5\ang$<$\ewoii$<10$\ang, compared to 74\% for our sample.  We can therefore say with confidence that our samples are largely disjoint and that they are focused on early-types with residual star formation, whereas we are focusing on galaxies with no recent star formation.  Nonetheless, their results have implications for gas processes in clusters, which we will address in subsequent sections.

\subsubsection{Could the pAGB properties be different in different environments?}
\label{Sec:pAGB_diff}

If we assume that the emission in our ``older" galaxies is indeed powered by pAGB stars, the next question is whether the pAGB properties could be different in different environments.   One possibility to explain the difference in pAGB properties would be if the stellar IMF were different in the various environments, as different abundances of low-mass stars would influence the number of UV-emitting evolved stars (pAGB or Horizontal Branch (HB) stars).  Indeed, \citet{Zaritsky14} and \citet{Zaritsky15} have shown that the total mass-to-light ratio of early-type galaxies drawn from the SAURON and ATLAS$^{\rm 3D}$ samples correlates strongly with UV color, implying that the low-mass IMF slope partially controls the UV emission properties in early-type galaxies.  They propose that the UV-slope could actually test for IMF differences.  Unfortunately, we do not have rest-frame UV photometry for our galaxies and can therefore not directly test if the IMF is different for galaxies in different environments.  However, there is no reason to assume that the galaxies in groups and clusters in our sample have different IMFs.  The mean velocity dispersions for cluster and field early-type galaxies in EDisCS above our stellar mass limit are within 0.05~dex of each other \citep{Saglia10}.  This implies at most a $\sim 0.03~$dex systematic  difference in the total mass-to-light ratios \citep{Zaritsky15}.  We therefore have no evidence that differences in the spectrum caused by IMF differences can explain the observed differences in the \ewoii\ distributions.

The only properties of the stellar population that could conceivably be different in different environments are the SFH and metallicity.  Regarding the latter, the mass range of our sample (log(\mstar/\msol$)>10.4$) makes it unlikely that there are substantial metallicity variations.  Regarding possible SFH differences, the number of pAGB stars in a stellar population, and hence their contribution to the UV radiation field, changes with stellar age.  \citet{Binette94} contains one of the most recent determinations of this contribution and finds that the UV continuum output of pAGB stars changes by $\sim 15\%$ for stellar populations between 1 and 8 Gyr, spanning the full possible range of age differences in our sample as governed by our \dn-\hdga\ selection and a formation redshift of 6.  However, these changes occur in an opposite sense shortward and longward of Lyman-$\alpha$, such that they mostly cancel out in terms of the total UV energy input \citep{Binette94}.

Despite the small expected difference in the pAGB flux, we nonetheless attempt to further constrain the potential age differences between different environments.  By virtue of their selection in \dn\ and \hdga\ strength, the ``older" galaxies in our different environmental bins are already constrained to have roughly similar ages.  They also have the same peak in their distributions in the \dn-\hdga\ plane (see Figure~\ref{Fig:d4hd_hist}), ruling out any large differences within this larger age bin.  Nonetheless there may be small age differences between passive galaxies in different environments.  Drawing from the literature,  \citet{vandokkum07} performed a fundamental plane comparison of visually classified massive E/S0 galaxies in the field and in rich clusters at $z<1$ and found that the luminosity weighted ages between these two populations differed by only 4\%.  This small age difference makes it unlikely that there could be large differences in the pAGB population driving the changes in the \oii\ fraction.  

On the other hand, \citet{Saglia10} measured the fundamental plane for the EDisCS clusters used in this analysis and found that the size at a fixed dynamical or stellar mass increases with decreasing redshift.  Taking this into account, and performing a consistent comparison between field and cluster galaxies they found that the field galaxies were approximately $1-2$Gyr younger than the cluster galaxies at a fixed stellar mass.  The origin of the difference between \citet{Saglia10} and \citet{vandokkum07} is not clear.  Perhaps it stems from the latter using much more massive clusters than the former as there is some tentative evidence for different speeds of the build up of the red sequence in low and high-mass clusters \citep{DeLucia07a,Gilbank08,Rudnick09}.  Additionally, \citet{vandokkum07} assumed homology, i.e. no size evolution in the early type population at a fixed surface brightness and velocity dispersion, whereas \citet{Saglia10} allowed for size evolution.  This may result in different estimates for the stellar population ages derived using the fundamental plan evolution.  We adopt the results from \citet{Saglia10} as an indication of the maximum age difference between cluster and field galaxies and explore the consequences below.  Unfortunately, the quality of the EDisCS data also precludes a precise estimate of the stellar population age differences between old galaxies with and without weak emission lines \citep{Sanchez09}.

It is not possible to easily convert this age difference into a difference in expected \oii\ fractions as the modeling of the UV contribution of pAGB stars is still very uncertain.  \citet{Brown08} find far fewer pAGB stars in M32 than expected by theoretical models, and similar in abundance to more recent observational studies of M31 by \citet{Rosenfield12}.  Other authors have pointed out that the pAGB contributions to the Lyman continuum of galaxies is uncertain at the factor of $\sim 2$ level \citep{Stasinska08,CidFernandes11,Papaderos13}.   However, we can investigate if this age difference can result in a significant difference in the number of young stars, under the assumption that they may contribute somewhat to the \oii\ excitation mechanism if they exist in small proportions.  According to Figure~\ref{Fig:d4hd}d, galaxies need at least six e-foldings of their SFH to enter the ``older" region of the \dn-\hdga\ plane.  For a $\tau=1$~Gyr exponential SFH (red curve in Figure~\ref{Fig:d4hd}d), the mass fraction of stars younger than 1~Gyr decreases from 0.5\% for a 6~Gyr old population to 0.2\% for a 7~Gyr old population.  While the fraction of these young stars for a simple tau model is very small, it is systematically different for field and cluster galaxies.  Unfortunately, it is quite complicated to convert this stellar population difference into a difference in the \ewoii\ distribution, especially as we do not know the relative spatial distribution of gas and young stars in our galaxies.  This renders us unable to rule out modest age differences contributing somewhat to the difference between field and cluster galaxies.  However, because the difference in \ewoii\ distributions is also present between groups and the field, it implies that systematic age differences must exist between those two environments if that were to explain our result.  Unfortunately, no comparisons of the fundamental plane evolution in groups and the field have been undertaken, making it impossible to determine if age differences could be explaining the difference in \oii\ fractions in those environments.  That said, as our groups may not exist in large-scale overdensities, it is not even clear if we would expect them to have collapsed at an earlier time than our field galaxies.

In summary, there may be small age differences for passive galaxies in different environments but there is no way to directly test the effect of these age differences on the \oii\ fractions.  Measuring the \oii\ fractions of galaxies in halos with a wider range in mass and large scale overdensity may help to solve this problem.  Additionally tracing the spatial distribution of the small young stellar populations and the gas may enable the use of models to determine if the \oii\ fraction difference can be caused by age differences.  Setting those uncertainties aside, we conclude that there are no required mechanisms for a difference in excitation source being responsible for the difference in \ewoii\ distributions.  

\subsection{The Environmental Dependence is not Driven by the Morphology Density relation.}
\label{Sec:morph_dep}

It is well known that the early-type fraction is higher in dense
environments than in the field \citep[e.g.][]{Dressler80b,Postman05}.
Therefore it is important to check that the difference in $f_{oe}$ are
not merely reflecting the environmentally dependent early-type
fraction.  We test this using the HST-based visual morphologies
available for 11 clusters and 5 groups in our sample.

\begin{figure}
\epsscale{1.00}
\plotone{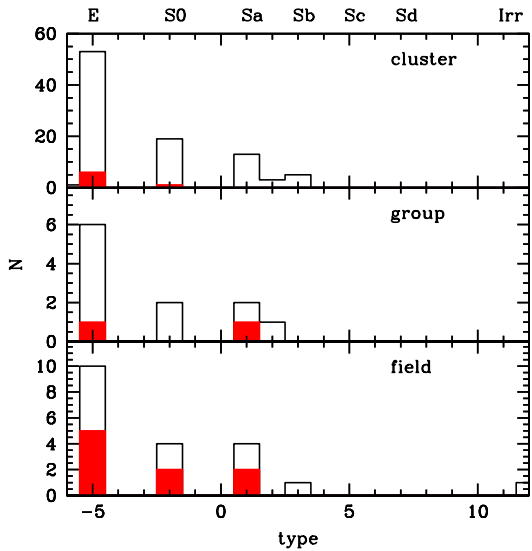}
\caption {The distribution of morphological types for the ``older''
  galaxies.  The T-Type is given on the bottom axis and the Hubble Type on the top axis.  The empty
  histogram is for all of these galaxies and the filled histogram is
  for the subset with \ewoii$>5$\ang.  The vertical scale is different
  in each panel.  All but three of the ``older'' \oii\ emitters have
  E/S0 morphology, indeed the early-type fraction of this subsample of
  galaxies is consistent at 1-$\sigma$ across all environments.  This
  indicates that the environmental difference in the
  \ewoii\ distribution of ``older'' galaxies cannot be driven by the
  morphology-density relation.}
\label{Fig:type_hist}
\end{figure}

In Figure~\ref{Fig:type_hist} we show the morphological distribution
of ``older'' galaxies in each environment.  76\% of our ``older''
galaxies are E/S0 and the remaining 24\% are mostly Sa with a few Sbs.
All but three of our emission-line ``older'' galaxies are early-type
galaxies.  The fraction of ``older'' galaxies that are early-types are
statistically indistinguishable in the different environments.  Also,
the fraction of the ``older'' emission-line galaxies that are early
types is $\sim 80\%$ and is statistically indistinguishable in the
different environments.  However, when restricting ourselves to
early-type galaxies, the difference in the fraction of emission-line
galaxies is as significant as for the whole spectroscopic sample.

We therefore conclude that changes in the morphological fraction among
our environments is not driving the difference we see in the
emission-line fraction.  

\subsection{The Origin of Differences in Gas Content}
\label{Sec:gasdif}

As outlined in the previous sections, the difference in the \ewoii\ distributions is neither due to differences in the radiation field, nor to differences in the morphological distribution of galaxies in different environments.  The remaining possibility is that the difference in the \ewoii\ distribution is caused by differences in the gas content.  In all galaxies gas is resupplied to the ISM via mass loss from stars.  For a Chabrier IMF, the total mass returned to the ISM from a simple stellar population reaches 50\% at an age of 6 Gyr \citep{Bruzual03}, which corresponds to the time difference between the lowest redshift end of our sample ($z=0.4$) and $z=1.7$.\footnote{By $z=1.7$ many red sequence galaxies were already
  in place \citep{Papovich10,Rudnick12}.}  in the first 1 and 2 Gyr, 80\% and 90\% of this gas is returned to the ISM respectively.  Therefore, the total gas content changes only by 10\% in the time between 2 and 6 Gyr, representing the largest possible difference in gas contents.  However, the age difference between the cluster and field galaxies in our sample is likely significantly less than this (see \S\ref{Sec:pAGB_diff}) and the difference in gas contents should therefore be correspondingly smaller.  It is therefore unlikely that differences in the amount of gas from mass loss resulting from age differences can account for the difference in the observed \oii\ fractions.  

In this subsection we therefore consider additional external origins for the gas difference and discuss whether the differences represent an excess of gas in field galaxies or an active removal of gas in cluster and group galaxies. 

\subsubsection{Excess of gas in field galaxies}
\label{Sec:field_excess}

A possible explanation for the difference in emission-line fractions in
old galaxies in different environments is that gas accretion onto
satellite galaxies in groups and clusters is no longer efficient.  In
this picture, the ``older'' emission-line galaxies that we see in
groups and clusters are also likely present in the field and have their gas supply
dominated by mass-loss from existing stellar populations heated by
evolved stellar populations.  Given that 90\% of the gas is returned in the first Gyr, there should be ample gas to supply this
emission.  The additional galaxies with
emission in the field would then come from a population of galaxies
that are undergoing accretion of gas directly onto those galaxies, in either a hot or cold mode of accretion.  Galaxies in the field with our stellar mass are likely to be central galaxies of their halos and so would be the repository of this gas, although it is presumably kept from forming stars by some sort of feedback.  Most of our galaxies in groups and clusters, however, are satellites and the gas that would normally have been deposited into their halos will instead presumably be deposited into the intragroup or intracluster medium in the more massive halos.  This lack of a direct accretion source onto the galaxy halo could then drive the difference in the gas contents.

It is possible that this gas is cooling directly onto the field
galaxies and not arriving in the form of infalling satellite galaxies.
Current theoretical pictures of galaxy evolution highlight the
importance of gas accretion in fueling star formation in galaxies.
This can either occur by the inflow of gas in cold filaments
\citep[e.g.][]{Keres05,Dekel09,Keres09} or by a more classical picture of gas
cooling onto a galaxy from a hot halo that consists of gas that was
shock-heated to the virial temperature during the infall process
\citep[e.g.][]{WhiteRees78}.  That the fraction of ``older'' galaxies
with emission is as low in groups as in clusters implies that merging
is not the main delivery mechanism for the gas as the lower velocity dispersions of the groups would imply a higher merger
cross-section than in clusters.  If minor mergers were the primary mechanism
for gas delivery then one would naively expect that the fraction of
``older'' galaxies with emission in groups would be as high as in the
field if not higher.  Supporting the idea that smooth gas accretion may be adding gas to field galaxies is the observational evidence from the COS-HALOS project for significant amounts of cool gas around low redshift early-type galaxies \citep{Thom12}, comparable to the amounts seen around star-forming galaxies.  This indicates that an ample circumgalactic medium built up by accretion from the cosmic web may be a common occurrence for passive galaxies.

This interpretation is consistent with studies of local early type
galaxies as part of the \atlas\ survey.  In \citet{Davis11}, $42\pm5\%$ of all field fast rotating early-types have ionized and molecular components that are kinematically misaligned with the stars, whereas misalignments are nearly absent for fast rotators in groups and in the Virgo cluster.  This argues for an accretion origin for much of the gas in the field, with the gas in clusters and groups having a purely internal origin thanks to the low accretion efficiency.  It was also shown that slow rotators exist only in the core of Virgo and always have misaligned ionized and molecular gas disks.  \citet{Davis11} interpreted the slow rotator gas properties as reflecting a different formation channel in which they formed by repeated major mergers \citep{Khochfar11}.  Indeed, from multiple studies of a $z=1.62$ cluster, there is evidence that mergers played an important role in the formation of massive red sequence galaxies \citep{Rudnick12, Papovich12, Lotz13}.  Since the location in the Virgo cluster core implies that the slow rotators are not freshly in fallen, the implication is that the misaligned gas must be able to survive the passage of the galaxy through the ICM, establishing a limit to the effectiveness of stripping from the centers of early types \citep[see \S\ref{Sec:dense_remove}][]{Davis11} although the molecular gas disks in Virgo early types do have smaller sizes than in the field \citep{Davis13}.  

Despite the support that the local studies lend to the accretion cutoff scenario, there are still difficulties in developing a consistent picture.  For example, it is puzzling that the detection rate of CO in Virgo galaxies is nearly constant with environment \citep{Young11},
in contrast to the HI detection rate \citep{Serra12}, implying that the physical
characteristics of the gas in the field and clusters may be different
\citep{Davis11,Davis13}.  Also, \atlas\ only observed the Virgo
cluster and so their cluster results may not be indicative of the
larger population at high density given the substantial cluster-to-cluster variance.

The implied prevention of gas cooling that we see in our group and
cluster galaxies is consistent with the phenomenological process
known as ``satellite quenching'', in which a galaxy's gas supply is
shut off if it becomes a subhalo of a more massive halo.  This process
is evident in the local universe in that
satellite galaxies at a fixed mass are found to have less activity
than centrals of the same mass \citep[e.g.][]{Pasquali09,Wetzel12}.  A
distinction between centrals and satellites frames our observed
phenomena in the context of dark matter halos and subhalos because in
our case every cluster and group galaxy except the BCG is by
definition a satellite galaxy, while in the field most of our
galaxies will be central galaxies of their own halos.  If we were to
interpret our observations in terms of the satellite/central
dichotomy, it would imply that such a quenching is effective even at
group scales with total masses of $10^{13}-10^{14}$\msol.  This halo mass range is consistent with other measures inferred from modeling the evolution and mass dependence in the passive fraction of galaxies \citep[e.g.][]{DeLucia12a}.

The mechanism for cutting off gas accretion likely involves a combination of tidal interactions and the ram pressure effects on the accreting gas.  The tidal effects can act both between the dark matter halo of the galaxy and the primary cluster or group that it is falling into as well as between the this primary halo and the gas that is in the circumgalactic medium or accreting onto galaxies from the cosmic web \citep{Hahn09}.  On the other hand, ram pressure stripping is canonically viewed as a process that only affects the ISM of a galaxy \citep[e.g.][]{Gunn72} but in reality ram pressure as a physical effect can also affect the
circumgalactic medium or infalling gas \citep{Bahe13}.  Both of
these can broadly be labeled strangulation or starvation as they do
not involve actively removing the gas from the depths of the potential
well (but see \S\ref{Sec:dense_remove}).  These mechanisms are thought to be effective
even at the group scales \citep{Kawata08}, lending support to our
picture.

We can attempt to understand this scenario in the context of the phase space distribution of our sources (\S\ref{Sec:phase}).  The old galaxies with emission are not found in the centers of the cluster in radius and velocity.  According to the models of \citet{Oman13}, this region is characterized by a long time since infall into the cluster, with very few galaxies in this region having entered the cluster in the last 1-2~Gyr.  It may therefore be that the galaxies with emission are those which have fallen in more recently, and thus are those which have had more time to accrete gas from the cosmic web or conversely less time to consume the gas that they had at the time of infall..  

An alternative but related mechanism to explain the different emission line fractions is if ``older" galaxies have a significant, though not necessarily dominant, formation channel through major mergers.  In the field, the gas ejected from a merger via tidal or feedback effects may fall back onto the galaxy after a few Gyr and replenish the gas supply.  In some cases this may rejuvenate star formation and regrow a disk \citep{Kannappan09,Moffet12} but depending on the specific characteristics of the merger, e.g. mass ratio, merger geometry, merger gas fraction, it may also just add a modest amount of gas to the galaxy.  While mergers may be more common in groups because of their low velocity dispersions and small distance between galaxies, the gas ejected during a merger in a dense environment is less likely to fall back onto the merged galaxy because of its interactions with the group tidal field and the intragroup medium.  This gas refueling mechanism is likely therefore more effective in field galaxies than in group or clusters and could explain some of the differences in gas content implied by our observations.  It is difficulty to quantify the prevalence of mergers in our sample as low surface brightness tidal features from mergers that happened a few Gyr in the past would be nearly impossible to detect. 

This gas refueling through merger fallback has a different origin than gas accretion through the cosmic web but has a similar effect on the gas contents and still represents a surplus in field galaxy gas supply compared to dense environments.  It is not yet clear if this scenario is consistent with the phase space distribution of galaxies presented in \S\ref{Sec:phase}.  On one hand, clusters are usually thought to have low merger fractions because of the high velocity dispersions.  On the other hand, 50\% our ``older" emission-line galaxies are outside the virialized region of phase space and may represent infalling galaxies for which the merger cross-section should be higher.  It is also unclear whether the ICM density and tidal forces in the cluster outskirts are sufficiently high to decouple the ejected merger gas.  Indeed, theoretical studies of the ram pressure in filaments and groups show that the ram pressure there can be elevated by $\sim 10-100$ compared to the field \citep{Bahe13} and so it is unlikely that merger ejecta would be able to efficiently reaccrete in cluster galaxies.  Therefore, it seems that this scenario is compatible with our observations.

In Figure~\ref{Fig:em_cumhist} we showed that \ewoii\ in the field was
shifted towards slightly higher values than in the clusters and
groups.  Since the continuum shapes (colors) of the ``older" galaxies (Fig.~\ref{Fig:colmass}) are very similar in different environments, this difference in \ewoii\ implies a differences in the flux of \oii.  While the significance of this shift is very low, one
possible interpretation is that the strength of the emission in the
different environments is different.  If we are sampling the tail of
the \ewoii\ distribution function, slightly lowering the mean
\ewoii\ of the sample would result both in a lower observed
\ewoii\ value and in a lower fraction.  All of the mechanisms we describe above could account for a decreased mean emission strength in cluster and group galaxies.  However, it is impossible to directly test
this assumption without more galaxies and significantly deeper
spectroscopy that would allow us to probe to much lower \ewoii.

The conclusion of this section is that a cutoff of accretion onto field galaxies seems to be required by the observations of galaxies in groups and clusters, given what we know about the reservoirs of cool gas around early type galaxies in the field, the effectiveness of strangulation in clusters, and the consistency with the phase space distribution of galaxies.  Merger fallback may be occurring in groups and is allowed by our observations, but can't explain the phase space distribution and so cannot be the dominant process at play.

\subsubsection{Removal of gas in cluster and group galaxies}
\label{Sec:dense_remove}

Another possibility to explain the observed differences in \ewoii\ distributions is that gas is being actively removed from within the optical radius of galaxies in dense environments.  This presumably would occur via ram pressure stripping of the gas by a galaxy's passage through the intracluster or intragroup medium.  Unfortunately, we only have ICM measurements for 3 of our systems \citep{Johnson06}.  The lack of X-ray data means that we cannot directly test the efficiency of ram pressure stripping for our systems.  All we can say is that our three X-ray observed clusters are consistent with the $L_x-T$ relation for X-ray selected clusters, so we can make the assumption that the stripping efficiency is not very different from the general cluster population.  

We attempt to place constraints on the role of stripping using the phase-space diagram shown in Figure~\ref{Fig:rad_v} and discussed in \S\ref{Sec:phase}.  It is clear that cluster galaxies with emission, regardless of their stellar age, and galaxies with young stellar populations, regardless of the presence of emission, are not found in the region of the cluster occupied by the old emission-line free galaxies.  One way to interpret this is as evidence in support of strong
ram-pressure stripping of the diffuse hot gas from the interiors of
the ``older'' massive cluster galaxy population.  In all other respects
the mass-selected ``older'' galaxies with and without emission are
identical and so the absence of the ``older'' emission-line galaxies
in the ''virialized'' regions of phase space must imply that the ionized gas emission is inhibited in
these objects.  In this scenario, the distribution of objects in Figure~\ref{Fig:rad_v} is consistent with a picture in which the
``older'' emission-line galaxies are infalling into the cluster for
the first time or that they happen to be on orbits that do not pass
close to the cluster cores.  This matches the simulations of \citet{Mahajan11} who find that between 48 and 83\% of galaxies lying at $0.5<R_{proj}/R_{200}<1.0$) and $1<|\Delta v|/\sigma<3$ are infalling for the first time. If stripping is occurring in the ``older" emission-line galaxies, our data are not sufficient for us to determine whether
the gas is fully stripped or just reduced enough so as to lower
\ewoii\ below our detection limit.  Spatially resolved and
high-sensitivity measurements of the the line emission will help to
resolve this situation.  

If we correct our ``older" emission-line fractions in clusters for the contamination of these infalling galaxies, it will remove proportionately more emission-line galaxies than passive galaxies as the latter are more centrally concentrated (Figure~\ref{Fig:rad_v_cumhist}).  Therefore the actual fraction of ``older" emission-line cluster galaxies we quote in this paper is likely an upper limit.

In studies of local galaxies from \atlas\ there is some evidence that gas in early-type galaxies can be directly modified by the environment.  \citet{Serra12} find that the HI content varies continuously with environment, being lowest in the center of the Virgo cluster and already reaching significant values by the outskirts of the cluster.  In addition, \citet{Davis13} find that  early-type galaxies
in Virgo have more compact CO profiles than those in the field and more commonly have
truncations or significant asymmetries than galaxies in the field
\citep{Davis11,Davis13}, implying a direct effect of the cluster environment on the gas properties of early-type galaxies.  However, \citet{Young11} find no significant difference in the molecular gas contents of \atlas\ early-type galaxies in different environments with only a slight suggestion that the most CO rich objects are in the lowest density environments.  This picture is further complicated by the presence of ionized and molecular gas in slow rotating early types residing in the core of the Virgo cluster, implying that this gas can remain in place despite the effects of the cluster environment \citep{Davis11}.  Clearly, the effect of environmental stripping on the gas in cluster early type galaxies is complicated and may involve multiple channels for supply and depletion.  We can therefore not rule out the role that stripping plays in modulating the gas supply in old early type galaxies in our clusters.

Given the low velocities of group galaxies relative to the group potential, it is unlikely that active removal of the gas is an important mechanism in group environments, although we cannot test that here.  

\section{Summary \& Conclusions}
\label{Sec:summ}

In this paper we perform a spectroscopic analysis of cluster, group,
and field galaxies at $0.4<z<0.8$ drawn from the ESO Distant Cluster
Survey (EDisCS).  We select galaxies with log(\mstar/\msol$)>10.4$ and
decompose their absorption and emission spectra using an iterative
fitting process.  From the decomposed spectra we then measure the
spectral indices \dn, \ewhda, \ewhga, \ewhbe, \ewoii~$\lambda3727$, and \ewoiii~$\lambda\lambda4959,5007$.  We use the
continuum and absorption indices to characterize the relative
luminosity-weighted mean age of the stellar populations for our
galaxies and use the emission indices to probe the state of the
ionized gas.  In this paper we have examined the emission-line
properties of stellar-mass selected galaxies in different environments but having the same
stellar age.

As has been found by many other authors, galaxies in all environments
lie on a locus in the \dn-\hdga\ plane and progress
towards older stellar ages in the direction of smaller H$\delta$ and H$\gamma$
absorption strength and larger \dn.  Using stellar population
models we define regions in the \dn-\hdga\ plane that
correspond to galaxies with generally ``older'', ``intermediate'', and
``younger'' stellar ages.  These regions can also be thought of as
marking regions of different numbers of star formation history
e-foldings.  Our continuum spectroscopic indices have the advantage
over broad-band colors in that they are more age sensitive and are
less susceptible to dust extinction.  For example, galaxies with
broad-band colors that would place them on the red sequence have a mix
of stellar ages, with both ``older'' and ``intermediate''-age stellar
populations.  There is a significant population of galaxies on the
blue side of the red sequence that have ongoing star formation and
whose red colors likely stem from dust extinction or from recently quenched star formation.

Our main observational results are as follows:  

\begin{itemize}

\item When considering only galaxies with old stellar ages we find that 8\%
and 6\% of the cluster and group galaxies respectively have weak
\oii\ emission.  In contrast 27\% of the analogously selected
``older'' field galaxies have this amount of emission.  The fractions
in the dense environments and the field are different at the
2.8$\sigma$ level.  None of the other age categories display a difference
in their emission-line fractions with environment.

\item The stellar mass distributions of the emission line galaxies in different environments are statistically consistent.   The difference in the fraction of
galaxies with emission in the different environments is true even at a
fixed stellar mass and is not a relic of the morphology density
relation.  We have also shown that
the observed environmental trend is not being driven by a few systems
with many members as every single cluster and group has an
emission-line fraction lower than that in the field.

\item We find that all galaxies with emission lines are less common in the central
regions of our clusters than ``older'' galaxies with no emission, which as a population
occupy the region of phase-space belonging to virialized
systems.  The subpopulation of ``older'' galaxies with emission are consistent with the phase space distribution of the other emission-line galaxies and therefore are also largely absent from the virialized regions of the cluster.

\end{itemize}

When H$\beta$ and \oiii\ emission are available we use them to diagnose the nature of the
emission.  We find that the line ratios of the emission in the
``younger'' and ``intermediate'' populations are consistent with normal
star formation.  For the ``older'' emission-line galaxies, however,
the emission is incompatible with star formation.  Drawing on the
many spatially-resolved spectroscopic studies of early-type galaxies
in the nearby Universe we conclude that this emission is likely not
from an AGN but is likely originating in gas that is heated by evolved
post-AGB stars.  

We regard all possibilities for the difference in the emission line fractions.  We rule out differences in the  radiation in different environments as driving the difference between field and cluster galaxies and conclude that it is driven by a difference in gas contents.  We then explore whether the difference in gas supply is driven by an extra source of gas in field galaxies or by the removal of gas in dense environments.  In all scenarios, galaxies are supplied by gas from stellar mass loss.  We conclude that field ``older'' galaxies are supplied by additional gas accretion from the cosmic web and that cluster and group galaxies have had their accretion shut off.  This conclusion is required by our observations as the deficit in the emission line fraction happens at group halo mass scales where active stripping is not thought to be active.  In addition, this scenario is consistent with the distribution of galaxies in phase space as galaxies nearest the virialized region also have the longest times since infall, and therefore may have had less time to accumulate gas before entering the cluster.  Our result that the emission-line
fractions are suppressed in halos of of mass $\sim
10^{13}$\msol\ implies that even groups are effective domains
of processes like  starvation.

An additional scenario is that a significant fraction of ``older'' galaxies are formed via mergers and that the gas falling back onto these galaxies significantly after the merger becomes decoupled in dense environments.  This is similar in spirit to the accretion cutoff scenario but involves gas from the original galaxies and not accretion from the cosmic web. This scenario is can be effective in both groups and infalling cluster galaxies but may not be effective for virialized cluster galaxies for which the merger cross-section is very low.

Finally, we consider active stripping of the gas via passage through the intracluster medium.  Local studies of the Virgo cluster from \atlas\ show that the cluster environment has an effect on the gas content of early-type galaxies, but one that is complex, with potentially multiple supply and depletion channels.  The lack of galaxies in the heart of the virialized region of our clusters is consistent with a picture in which gas is stripped out of the galaxies. This scenario is clearly allowed, though not required, by our data.  Additionally, because the emission line deficit happens even at group mass scales where stripping is not thought to be effective, it cannot be the sole process in modulating the gas contents of passive galaxies.  To more conclusively determine the different channels of gas depletion will require deep spectroscopic observations of a wider range of environments, and for enough galaxies to tease out differences between clusters, groups, and the field.  Having high spatial and spectral resolution observations that would help us constraint the geometry and kinematics of the gas would be especially beneficial.




\acknowledgments

GHR acknowledges funding support from HST program HST-GO-12590.011-A, HST-AR-12152.01-A, HST-AR-14310.001, NSF AST grants 1211358 and 1517815, and the NSF under Award No. EPS-0903806 with matching support
from the State of Kansas through Kansas Technology Enterprise
Corporation.  GHR would also like to acknowledge the support of an Alexander von Humboldt foundation fellowship for experienced researchers and the excellent hospitality of the Max-Planck-Institute for Astronomy, the University of Hamburg Observatory, the Max-Planck-Institute for Extraterrestrial Physics, the International Space Sciences Institute, and the European Southern Observatory, where some of this research was conducted.  GHR would 
like to thank Tracy Webb and Alison Noble for useful discussions
regarding the use of phase-space diagrams to disentangle the accretion
histories of galaxies, Glenn van der Wenn, Marc Sarzi, and Bernd Husemann for helping to compare
our results to those from local studies of early-types, and Jason Spyromilio for discussions about the physical meaning of equivalent width measurements.  
YJ acknowledges support by FONDECYT grant No. 3130476. YJ also acknowledges support from the Marie Curie Actions of the European Commission (FP7-COFUND). BMJ acknowledges support from the ERC-StG grant EGGS-278202.
The Dark Cosmology Centre is funded by the DNRF.

\bibliographystyle{aasjournal}
\bibliography{references}

\clearpage

\LongTables
\setcounter{table}{1}
\begin{landscape}
\begin{deluxetable}{clllllllll}
\tablecaption{Galaxy Data}
\tablewidth{0pt}
\tablehead{\colhead{ID} & \colhead{$z_{spec}$} & \colhead{\ewoii\tablenotemark{a}} & \colhead{\ewoiii\tablenotemark{b}} &  \colhead{\ewhbe}  & \colhead{\ewhda} &  \colhead{\ewhga} & \colhead{$D_n$(4000)} & \colhead{$\log(M_*/M_\odot)$} & \colhead{Environment}\\
\colhead{} & \colhead{} & \colhead{[\ang]} &  \colhead{[\ang]}  & \colhead{[\ang]} &  \colhead{[\ang]} & \colhead{} & \colhead{} & \colhead{}}
\startdata
EDCSNJ1018471-1210513 & $0.4716$ & $  6.64\pm  1.20$ & ... & $  1.42\pm  0.33$ & $  1.37\pm  0.60$ & $ -1.57\pm  0.61$ & $  1.68\pm  0.03$ & $ 10.97$
 & cluster\\
EDCSNJ1018464-1211205 & $0.4717$ & $<  0.77$ & ... & ... & $ -0.05\pm  1.03$ & $ -4.93\pm  1.08$ & $  1.93\pm  0.07$ & $ 10.61$
 & cluster\\
EDCSNJ1018467-1211527 & $0.4716$ & $<  0.51$ & ... & ... & $ -1.13\pm  0.37$ & $ -5.55\pm  0.37$ & $  1.96\pm  0.03$ & $ 11.63$
 & cluster\\
EDCSNJ1018454-1212235 & $0.4789$ & $  6.48\pm  0.66$ & $  1.34\pm  0.36$ & $  3.59\pm  0.25$ & $  2.82\pm  0.43$ & $  1.29\pm  0.38$ & $  1.31\pm  0.02$ & $ 10.94$
 & cluster\\
EDCSNJ1018438-1212352 & $0.4732$ & $<  1.10$ & ... & ... & $  2.60\pm  0.90$ & $ -2.50\pm  0.90$ & $  1.67\pm  0.05$ & $ 10.52$
 & cluster\\
EDCSNJ1018483-1212474 & $0.4727$ & $  5.64\pm  0.66$ & $  1.54\pm  0.32$ & $  4.09\pm  0.24$ & $  2.87\pm  0.44$ & $  1.65\pm  0.46$ & $  1.39\pm  0.02$ & $ 10.87$
 & cluster\\
EDCSNJ1018490-1212553 & $0.4733$ & $ 10.95\pm  0.66$ & $  2.47\pm  0.35$ & $  6.74\pm  0.22$ & $  5.49\pm  0.45$ & $  3.07\pm  0.48$ & $  1.30\pm  0.02$ & $ 10.77$
 & cluster\\
EDCSNJ1018473-1213164 & $0.4756$ & $ 10.58\pm  1.05$ & ... & $  3.64\pm  0.31$ & $  3.64\pm  0.63$ & $  2.38\pm  0.55$ & $  1.41\pm  0.03$ & $ 10.64$
 & cluster\\
\enddata
\label{indices_tab}
\tablenotetext{a}{\ion{O}{2}]~$\lambda3727$}
\tablenotetext{b}{[\ion{O}{3}]~$\lambda\lambda4959,5007$}
\tablecomments{All galaxies with \ewoii$<5$\ang\ have been replaced by their 1$-\sigma$ upper limits.  All galaxies with \ewhbe\ measurements that are less than 2$-\sigma$ are replaced by the 1$-\sigma$ upper limits. No entry for a given line indicates that the specified feature did not fall in the wavelength range of the spectrum.}
\end{deluxetable}
\newpage
\end{landscape}

\end{document}